\documentclass[a4paper,12pt]{article}
\usepackage{jheppub} 
\usepackage{lineno}
\usepackage{float}

\def\dgvv0{\delta g_{vv}^{(0)}}
\def\dgxv0{\delta g_{xv}^{(0)}}
\def\dgxx0{\delta g_{xx}^{(0)}}
\def\dphi0{\delta \phi^{(0)}}

\def\V{V(\Phi_0)}
\def\Vp{V'(\Phi_0)}
\def\Vpp{V''(\Phi_0)}

\def\lb{\bar{\lambda}}

\title{Pole skipping from universal hydrodynamics of (1+1)d QFTs}

\author{Richard A.~Davison$^{1}$ and Hanzhi Jiang$^{2}$} 

\affiliation{ $^{1}$ Department of Mathematics and Maxwell Institute for Mathematical Sciences, Heriot-Watt University,
Edinburgh EH14 4AS, U.K.}

\affiliation{ $^{2}$ Rudolf Peierls Centre for Theoretical Physics, University of Oxford, 1 Keble Road, Oxford, OX1 3PU, U.K.}

\emailAdd{r.davison@hw.ac.uk}
\emailAdd{hanzhi.jiang@physics.ox.ac.uk}       

\abstract{(1+1)d QFTs provide a tractable arena for understanding the emergence of hydrodynamics in thermal states. At high temperatures this process is governed by the weak breaking of conformal symmetry, and so in this limit many features of the hydrodynamic theory that emerges have been argued to be universal. In this paper we study aspects of the stress tensor thermal two-point function in holographic QFTs of this kind and show that they are consistent with the universal hydrodynamic theory proposed to apply at late times. Specifically, we identify the locations of the `pole skipping' points in momentum space at which there is an intersection of poles and zeroes of this two-point function in holographic QFTs. Although these points lie outside the regime where the hydrodynamic theory is controlled, we show that their locations are consistent with those found by resumming the hydrodynamic derivative expansion near the lightcone. For example, this resummation of the universal hydrodynamics correctly predicts the butterfly velocity of holographic theories.} 

\begin{document}
\maketitle
\flushbottom

\section{Introduction}
\label{sec:intro}

\paragraph{}A generic interacting QFT is expected to thermalise. One of the key steps in a typical thermalising system is the emergence of dissipative hydrodynamics at late times. This occurs when the system has equilibrated locally, and the subsequent relaxation back to global equilibrium is described by an effective theory for a fluid-like state. An important question in understanding thermalisation is therefore to determine how exactly hydrodynamics emerges from the microscopic dynamics of a given QFT. In particular, answering this would tell us how the properties of the macroscopic fluid that forms are related to the underlying microscopic constituents and their interactions. This is challenging, even in weakly coupled QFTs \cite{Jeon:1994if,Jeon:1995zm}.

\paragraph{}Restricting to (1+1) dimensions provides a simplified setting for addressing this topic. In (1+1)d CFTs, conformal symmetry prevents the emergence of dissipative hydrodynamics. Therefore its emergence in generic (1+1)d QFTs is tied to the breaking of this symmetry. We consider (1+1)d QFTs obtained by deforming CFTs with a relevant scalar primary operator $\mathcal{O}$ with dimension $0<\Delta<2$
\begin{equation}
\label{eq:deformedaction}
	S=S_{\text{CFT}}+\sqrt{c}\lambda\int dtdx\, \mathcal{O},
\end{equation}
where $c$ is the central charge of the CFT. Conformal symmetry is weakly broken at high temperatures as the dimensionless coupling in units of temperature $\bar{\lambda}=\lambda/T^{2-\Delta}$ is small in this limit. Conformal perturbation theory (i.e.~a small $\bar{\lambda}$ expansion) then provides a theoretical tool with the potential to address the emergence of hydrodynamics for QFTs at high temperature.

\paragraph{}A small symmetry-breaking parameter $\bar{\lambda}\ll1$ is expected to lead to dissipative hydrodynamics emerging at late times $t_{\text{eq}}\sim\bar{\lambda}^{-2}/T$ \cite{Delacretaz:2021ufg}. However, this emergence is reflected in the breakdown of conformal perturbation theory for thermal correlators near the lightcone at late times $t\sim t_{\text{eq}}$ \cite{Davison:2024msq}. In other words, accessing the hydrodynamic regime requires resumming conformal perturbation theory, even when $\bar{\lambda}\ll1$.\footnote{While this is unfortunate, it is still better than the situation in higher dimensions where for a generic QFT the emergence of hydrodynamics is not related to any small parameter at all.}

\paragraph{}Nevertheless, a proposal was recently made for the stress tensor thermal two-point function in the hydrodynamic limit for theories with large $c$ and $\bar{\lambda}\ll1$ \cite{Davison:2024msq}. The fluid-like state that emerges at late times is characterised by transport coefficients (speed of sound, viscosity, etc.) and infinitely many of these appear in the stress tensor thermal two-point function. The proposal of \cite{Davison:2024msq} provides an explicit relation for every one of these transport coefficients, to order $\bar{\lambda}^2$, in terms of the microscopic action \eqref{eq:deformedaction}. In fact, these relations are universal: they depend only on the dimension $\Delta$ and are otherwise completely independent of the original CFT.

\paragraph{}A complementary approach to understanding thermalisation in (1+1)d QFTs is to examine specific theories where direct calculations are possible. Holographic theories provide an excellent arena for this: in the large $c$ limit their thermal correlators can in principle be calculated numerically for any $\bar{\lambda}$, avoiding the issues described above that plague real-time perturbation theory. Even better, there are certain non-equilibrium QFT observables that are naturally geometrised in holographic theories such that a controlled perturbative expansion in $\bar{\lambda}$ can be easily implemented. Specifically, these are quantities that can be expressed directly in terms of the spacetime geometry characterising the equilibrium state.\footnote{A well-known example in higher dimensions is the shear viscosity: while fundamentally a non-equilibrium observable, in holographic theories it is directly related to the entropy density of the equilibrium state \cite{Kovtun:2004de}.} For these, the expansion in $\bar{\lambda}$ can be implemented directly on the equilibrium geometry: being time-independent, it does not suffer from the late time breakdown described above. In \cite{Davison:2024msq} this approach was used to show that the bulk viscosity of holographic theories at small $\bar{\lambda}$ \cite{Yarom:2009mw} agrees with the proposed universal expression.

 \paragraph{}In this paper we extend the study of non-equilibrium properties of these holographic theories in the high temperature limit $\bar{\lambda}\ll1$. Specifically, we investigate pole skipping in the retarded thermal two-point functions of the stress tensor in momentum space $G(\omega,k)$. Pole skipping refers to the phenomenon where, at an isolated set of points $(\omega,k)$ in (complexified) momentum space, $G(\omega,k)$ is undefined due to the intersection of a pole and a zero. While a seemingly abstract feature of the correlator, the location of pole skipping points explicitly contains information about fundamental thermalisation properties of the state \cite{Grozdanov:2017ajz,Blake:2017ris,Blake:2018leo}. Furthermore, in holographic theories the locations of these points can be related directly to the equilibrium geometry \cite{Blake:2018leo,Blake:2019otz} and therefore computed relatively easily at small $\bar{\lambda}$. This makes them an excellent testing ground for the universal hydrodynamics proposed in \cite{Davison:2024msq}.

\paragraph{}In holographic theories, pole skipping points exist when the corresponding graviton modes exhibit an extra solution that is ingoing at the black hole horizon.\footnote{Although in pure AdS$_3$ gravity all solutions are (large) gauge transformations, when $\lambda\ne0$ there are also non-trivial solutions due to the coupling of the graviton to a scalar field. This is the gravitational origin of the emergence of dissipative hydrodynamics.} In holographic QFTs in (2+1) (and higher) dimensions this happens generally \cite{Blake:2018leo} for the frequency $\omega=\omega_*$ and wavenumbers $k^2=k_*^2$ where\footnote{See \cite{Grozdanov:2018kkt,Li:2019bgc,Ahn:2019rnq,Abbasi:2019rhy,Liu:2020yaf,Ahn:2020bks,Sil:2020jhr,Blake:2021hjj,Mahish:2022xjz,Wang:2022mcq,Amano:2022mlu,Grozdanov:2023txs,Wang:2024byb,Ahn:2025exp,Lyu:2025kep} for generalisations of \cite{Blake:2018leo} to other spacetimes.}
\begin{equation}
\label{eq:vBintro}
    \omega_*=+i2\pi T,\quad\quad\quad\text{and}\quad\quad\quad k_*^2=-\left(\frac{2\pi T}{v_B}\right)^2.
\end{equation}
The butterfly velocity $v_B$ is a fundamental speed that characterises out-of-time-ordered correlations of local operators in the thermal state \cite{Roberts:2014isa}, and its appearance in $k_*$ is evidence that these correlations are governed by a simple effective theory for scrambling \cite{Blake:2017ris,Blake:2021wqj} (see also \cite{Knysh:2024asf} for further connections between this effective theory for scrambling and gravity). In holographic theories, $v_B$ is also the speed characterising operator growth in the thermal state, as measured by the bulk entanglement wedge \cite{Mezei:2016wfz,Dong:2022ucb,Baishya:2024gih,Chua:2025vig} (see also \cite{Lilani:2025wnd,Basu:2025exh,Deng:2025plg}). There are also generically pole skipping points for $\omega=\omega_n=-i2\pi T n$ ($n=0,1,2,\ldots$) and appropriately chosen $k^2=k_n^2$ \cite{Blake:2019otz} (see also \cite{Grozdanov:2019uhi,Natsuume:2019xcy,Natsuume:2019vcv,Wu:2019esr,Natsuume:2020snz,Ning:2023ggs}). 

\paragraph{}We show that these general results continue to hold for holographic theories dual to QFTs with actions of the form \eqref{eq:deformedaction}. By determining the equilibrium spacetime to quadratic order in small $\bar{\lambda}$, we obtain explicit, closed-form expressions for $v_B$, $k_*$, and $k_n$ (up to $n=2$) to this order in a high temperature expansion. The expressions for $k_*$ (and therefore $v_B$) and one of the $k_1$ are universal in that they depend only on the value of $\Delta$. When $\Delta=3/2$ and $\Delta=1,2$ our expression for $v_B$ reduces to that in \cite{Jiang:2024tdj} and \cite{Asplund:2025nkw}, and for general $\Delta$ it agrees with the recent numerical results in \cite{Asplund:2025nkw}. The expressions for the other $k_1$ as well as $k_n$ for $n\geq 2$ are non-universal in that they are sensitive to OPE coefficients of the operator $\mathcal{O}$, as well as its dimension.

\paragraph{}More importantly, we explain how these results can also be obtained directly from the proposed universal hydrodynamics of \cite{Davison:2024msq}. This is a little subtle. The hydrodynamic theory provides dispersion relations $\omega(k)$ of poles and zeroes that are expressed as series in $k$, with each coefficient determined to order $\bar{\lambda}^2$. However, the locations of the pole skipping points lie outside the radius of convergence of these series: they are sensitive to what happens prior to the emergence of hydrodynamics. To access this regime, we take the high temperature, near-lightcone limit
\begin{equation}
\label{eq:IntroHighTNearLCLimit}
    \omega\pm k\sim\bar{\lambda}^2\ll 1,
\end{equation}
of the proposed universal hydrodynamics of \cite{Davison:2024msq} and then resum in $k$ the stress tensor thermal-two point function (see the closed-form expression \eqref{eq:resummedG} and \eqref{eq:resummedGamma} below). By extending the regime of validity of universal hydrodynamics near the lightcone to early times in this way, we find an expression for $k_*$ (and therefore $v_B$) and one of the $k_1$ that agree with those found holographically. 

\paragraph{}This is very non-trivial evidence that the universal hydrodynamics proposed in \cite{Davison:2024msq} is indeed what emerges at late times in holographic theories. In fact it suggests that, after the resummation just described, the proposed universal hydrodynamics provides a good description of the dynamics near the lightcone at all times. Due to this, it directly contains information about early time scrambling such as $v_B$. The locations of the other $k_1$ as well as $k_n$ for $n\geq 2$ cannot be obtained from hydrodynamics in this way as they are far from the lightcone.

\paragraph{}We present our results in the opposite order from that described above. In Section \ref{sec:hydroreview} we briefly review the origin and structure of the hydrodynamic theory proposed in \cite{Davison:2024msq} and explain how to resum it in the high temperature, near-lightcone limit to obtain expressions for pole skipping locations and the butterfly velocity. In Section \ref{sec:Equilibriumsolution} we switch to holographic QFTs, perturbatively construct the equilibrium states at high temperature, and show that $v_B$ agrees with that just predicted. In Section \ref{sec:PoleSkipping} we determine the locations of pole skipping points in holographic theories dual to (1+1)d QFTs. We show that in the high temperature limit, those near the lightcone are consistent with resummed universal hydrodynamics, while those away from the lightcone are non-universal. We close in Section \ref{sec:discussion} with a discussion of the significance of the resummed two-point function.

\paragraph{}As this work was nearing completion the paper \cite{Asplund:2025nkw} appeared, in which the location of one of the pole-skipping points $(\omega_*,k_*)$ was extracted directly from first order conformal perturbation theory for the cases $\Delta=1,2$. Our expressions for this point in these cases agree. In Appendix \ref{app:comparison} we make a more detailed comparison. Our high temperature, near-lightcone resummation has consistency with first order conformal perturbation theory built-in, and we believe this is the appropriate way to understand this result.

\section{Butterfly velocity and pole skipping from hydrodynamics}
\label{sec:hydroreview}

\paragraph{}In this Section we will first briefly review the hydrodynamic theory of \cite{Davison:2024msq}. This is proposed to apply to (1+1)d QFTs in the limit $c\rightarrow\infty$ and at high temperatures $\bar{\lambda}\ll1$, with all transport coefficients that appear in the thermal two-point function of the stress tensor dependent only on $\Delta$. We then show how to extend this theory to shorter scales by taking the near-lightcone limit $\omega\pm k\sim\bar{\lambda}^2$ of the hydrodynamic two-point function and subsequently resumming in $k$. This yields predictions for the locations of pole skipping points, and the butterfly velocity, of holographic theories of this kind.

\subsection{Review of universal (1+1)d hydrodynamics}

\paragraph{}In a 2D CFT, the thermal two-point functions of the stress tensor $G$ are fixed entirely by conformal symmetry. When conformal symmetry is weakly broken, we can compute corrections perturbatively in the dimensionless coupling $\bar{\lambda}$
\begin{equation}
\label{eq:naiveCPTstruc}
    G=G_{\text{CFT}}+\bar{\lambda}^2G_{2}+\ldots.
\end{equation}
In this conformal perturbation theory, $G_2$ and subsequent corrections can in principle be computed from CFT correlators. The explicit results for $G_2$ are reviewed in Appendix \ref{app:comparison}. This expansion would seem to be useful for small $\bar{\lambda}$ (i.e.~sufficiently high temperatures). However, even when $\bar{\lambda}$ is very small this expansion is expected to break down at late times $Tt\sim\bar{\lambda}^{-2}$ due to the emergence of dissipative hydrodynamics.

\paragraph{}To obtain the correct late-time dynamics in the non-conformal theory, the naive conformal perturbation theory expansion in $\bar{\lambda}$ \eqref{eq:naiveCPTstruc} must be resummed. This can be seen very explicitly by working in momentum space. For definiteness we will take $G$ to be the thermal retarded two-point function of the energy density\footnote{All other two-points of the stress tensor are related directly to this by Ward identities: see Appendix \ref{app:comparison} for explicit expressions.} 
\begin{equation}
\label{eq:CPTstructure}
    G(\omega,k)=-\frac{c}{12\pi}k^2\frac{(2\pi T)^2+k^2}{\omega^2-k^2}+\frac{\pi c}{6}T^2+\bar{\lambda}^2G_2(\omega,k)+\ldots,
\end{equation}
where the first two terms on the right hand side are the thermal CFT result and $G_2(\omega,k)$ is given explicitly in equation \eqref{eq:G2correctionapp} below. Hydrodynamics is an effective theory which fixes the structure of this two-point function once it has emerged. Specifically, when the CFT central charge $c$ is large
\begin{equation}
\label{eq:HydroCorrelator}
G_{\text{hydro}}(\omega,k)=-k^2\frac{(\varepsilon+P)+\frac{c}{12\pi}k^2\kappa(\omega,k^2)}{\omega^2-c_s^2k^2-i\omega k^2\Omega(\omega,k^2)}+\varepsilon, 
\end{equation}
where $\varepsilon(\bar{\lambda})$ and $P(\bar{\lambda})$ are the thermal expectation values of the stress tensor components $T^{tt}$ and $T^{xx}$, $c_s^2=dP/d\varepsilon$, and $\Omega(\omega,k^2)$ and $\kappa(\omega,k^2)$ are infinite series of the form
\begin{equation}
\label{eq:transportexpansions}
	\begin{aligned}     &\,\Omega(\omega,k^2)=\Omega_1(\bar{\lambda})-i\omega\Omega_2(\bar{\lambda})-k^2\Omega_3(\bar{\lambda})+i\omega k^2\Omega_4(\bar{\lambda})+\ldots,\\
&\,\kappa(\omega,k^2)=\kappa_{2,0}(\bar{\lambda})-i\omega\kappa_{3,0}(\bar{\lambda})-\omega^2\kappa_{4,0}(\bar{\lambda})-k^2\kappa_{4,1}(\bar{\lambda})+\ldots.
	\end{aligned}
\end{equation}
The transport coefficients $\Omega_n(\bar{\lambda})$ and $\kappa_{n,m}(\bar{\lambda})$ play the role of the coupling constants of the effective theory: in principle they depend on the details of the underlying CFT and the choice of symmetry-breaking operator $\mathcal{O}$. Conceptually hydrodynamics is expected to be valid when $\omega$ and $k$ are sufficiently small: this is reflected in the series expansions \eqref{eq:transportexpansions} which follow from a derivative expansion in the real space formulation of the effective theory. The precise range of validity depends on the values of the transport coefficients in the series \eqref{eq:transportexpansions} and therefore on the details of the underlying CFT and choice of $\mathcal{O}$.

\paragraph{}It is clear that to obtain the hydrodynamic result \eqref{eq:HydroCorrelator} from conformal perturbation theory \eqref{eq:CPTstructure}, resummation in $\bar{\lambda}$ is required. In particular, this is necessary to generate the crucial $\bar{\lambda}$-dependent contributions $\Omega_n$ to the denominator of the two-point function. In a thermal CFT the energy density propagates freely at the speed of light, and it is these contributions that result in the energy density instead spreading and decaying as it is carried by dissipative hydrodynamic sound waves through the fluid.

\paragraph{}The above results are fixed on general grounds by symmetries. A proposal was made in \cite{Davison:2024msq} for how to also explicitly obtain the leading $\bar{\lambda}^2$ dependence of all transport coefficients $\Omega_n(\bar{\lambda})$ and $\kappa_{n,m}(\bar{\lambda})$. Rather than directly resumming conformal perturbation theory, this argument relied on assuming commutation of the small $\bar{\lambda}$ and hydrodynamic (small $\omega,k$) limits  in the two-point function of the trace of the stress tensor. Taking first the hydrodynamic limit, the effective theory fixes the structure of this object in terms of the transport coefficients, analogously to \eqref{eq:HydroCorrelator}. On the other hand, by taking first the small $\bar{\lambda}$ limit, we can evaluate this object explicitly using conformal perturbation theory. Comparing these in their overlapping regime of validity then gives explicit expressions for all transport coefficients at leading order in $\bar{\lambda}^2$. Specifically, 
\begin{equation}
\label{eq:generatingfunctional}
\begin{aligned}
	((2\pi T)^2+k^2)(&\,1-c_s^2-i\omega\Omega(\omega,k^2))-(\omega^2-k^2)\left(\kappa(\omega,k^2)-1\right)\\
	&\,=12\pi\lambda^2(2-\Delta)^2\left(G^{\text{CFT}}_{\mathcal{O}\mathcal{O}}(\omega,k)-\frac{\Delta}{(2-\Delta)}G^{\text{CFT}}_{\mathcal{O}\mathcal{O}}(0,0)\right),
	\end{aligned}
\end{equation}
where $G^{\text{CFT}}_{\mathcal{O}\mathcal{O}}(\omega,k)$ is the thermal retarded two-point function of $\mathcal{O}$ in the CFT, and the right hand side should be interpreted as a series in $\omega,k$. $G^{\text{CFT}}_{\mathcal{O}\mathcal{O}}(\omega,k)$ appears on the right hand side of this expression as, due to the dilatation Ward identity, it controls the first correction to the two-point function of the trace of the stress tensor in conformal perturbation theory. Not only is the form of $G^{\text{CFT}}_{\mathcal{O}\mathcal{O}}(\omega,k)$  universal -- it depends only on the dimension $\Delta$ -- but it is known explicitly \cite{Sachdev_1994}
\begin{equation}
\label{eq:OOCFT}
	G^{\text{CFT}}_{\mathcal{O}\mathcal{O}}(\omega,k)=\pi\left(2\pi T\right)^{2(\Delta-1)}\frac{\Gamma(1-\Delta)\Gamma\left(\frac{\Delta}{2}-\frac{i(\omega+k)}{4\pi T}\right)\Gamma\left(\frac{\Delta}{2}-\frac{i(\omega-k)}{4\pi T}\right)}{\Gamma(\Delta)\Gamma\left(1-\frac{\Delta}{2}-\frac{i(\omega+k)}{4\pi T}\right)\Gamma\left(1-\frac{\Delta}{2}-\frac{i(\omega-k)}{4\pi T}\right)}.
\end{equation}
As a consequence, equation \eqref{eq:generatingfunctional} yields explicit expressions for all transport coefficients at leading order in $\bar{\lambda}^2$ that depend only on $\Delta$ and are otherwise independent of the details of the QFT. In this sense the hydrodynamics is universal at high temperatures. There is a conceptual way to understand the simplicity of this result. At all times the CFT stress tensor is governed by a trivial, non-dissipative hydrodynamics where almost all transport coefficients are zero. Weakly breaking conformal symmetry then generates small values of all transport coefficients which, at leading order, are governed by the correlator of the symmetry-breaking operator $\mathcal{O}$ in the symmetric state.

\paragraph{}The explicit expressions for the transport coefficients at small $\bar{\lambda}$ allow us to answer more precisely the question of when the hydrodynamic expression \eqref{eq:HydroCorrelator} for the two-point function is valid. A natural way to identify this is through the dispersion relations $\omega_\pm(k)$ of its poles, which characterise the fundamental excitations of the thermal state. The hydrodynamic theory yields $\omega_\pm(k)$ as Taylor series in $k$, and the radius of convergence of these series defines a length scale below which the effective theory breaks down \cite{Withers:2018srf,Grozdanov:2019kge,Grozdanov:2019uhi,Heller:2020uuy}. After extracting the transport coefficients from \eqref{eq:generatingfunctional}, the high temperature hydrodynamic dispersion relations may be written
\begin{equation}
\begin{aligned}
\label{eq:resummedhydrodispersion}
    	\omega_{\pm}(k)=\pm k\left(1-\lambda^2\frac{6\pi(2-\Delta)^2}{(2\pi T)^2+k^2}\left(G^{\text{CFT}}_{\mathcal{O}\mathcal{O}}(\pm k,k)-\frac{\Delta}{(2-\Delta)}G^{\text{CFT}}_{\mathcal{O}\mathcal{O}}(0,0)\right) + \ldots\right), 
   \end{aligned}
\end{equation}
where $\ldots$ denotes corrections that are higher order in $\bar{\lambda}$. As it is a prediction of the hydrodynamic theory, the right hand side of this should be understood as a series in $k$. However, in the form written this expansion has been resummed.\footnote{This resummation in $k$ is different than the resummation of conformal perturbation theory in $\bar{\lambda}$ discussed around equation \eqref{eq:CPTstructure} above.} The resummation makes it easy to identify the radius of convergence as $k_{\text{eq}}=\Delta\pi T$: this is set by a pole in $G^{\text{CFT}}_{\mathcal{O}\mathcal{O}}(\pm k,k)$ and corresponds physically to the wavevector at which non-hydrodynamic thermal excitations of $\mathcal{O}$ in the CFT have a lifetime equal to that of the hydrodynamic sound waves. This pole is expected to be resolved into a branch point singularity when the full $\bar{\lambda}$ dependence of the dispersion relations is taken into account.\footnote{We emphasise that here we are referring to a pole in the dispersion relation rather than a pole in the correlator.}

\subsection{Near-lightcone resummation and pole skipping locations}

\paragraph{}We are now going to use this hydrodynamic theory to predict the locations of the pole skipping points in momentum space, and the butterfly velocity $v_B$, for holographic QFTs of this type. This will ultimately require taking a near-lightcone limit and then resumming the hydrodynamic derivative expansion.

\paragraph{}We begin by reviewing pole skipping in a CFT ($\lambda=0$). The stress tensor two-point function in this case is given by the first two terms in \eqref{eq:CPTstructure}. The pole skipping points are the $(\omega,k)$ for which there is an intersection of a pole and a zero of this correlator. Setting both the numerator and the denominator in the first term to zero separately\footnote{The second term does not affect the locations of pole skipping points as it is analytic in $\omega,k$.} yields solutions at three different frequencies: one in the upper half-plane at $\omega_*=+i2\pi T$, $k_*^2=-(2\pi T)^2$, one at the origin $\omega_0=0$, $k_0=0$, and one in the lower half-plane at $\omega_1=-i2\pi T$, $k_1^2=-(2\pi T)^2$ \cite{Haehl:2018izb}. The upper half-plane pole skipping point $\omega_*$ is particularly important: in maximally chaotic theories, it is conjectured \cite{Blake:2017ris} (and proven for many theories with gravitational duals \cite{Blake:2018leo}) that there is always a pole skipping point at this frequency and $k_*^2=-(2\pi T/v_B)^2$, where $v_B$ is the butterfly velocity that characterises the propagation of out-of-time-ordered correlations in the state. For a maximally chaotic (1+1)d CFT this therefore predicts $v_B=1$, which agrees with an explicit computation of the out-of-time-ordered correlator \cite{Roberts:2014ifa}. See \cite{Das:2019tga,Ramirez:2020qer} for generalisations to (1+1)d CFTs on other manifolds.

\paragraph{}When conformal symmetry is broken, the dispersion relations of the poles and zeroes will now depend on $\bar{\lambda}$ and thus so will the locations of the pole skipping points. Even when $\bar{\lambda}\ll1$, conformal perturbation theory \eqref{eq:CPTstructure} is no use on its own: without resummation in $\bar{\lambda}$ it contains no information on how pole locations depend on $\bar{\lambda}$. Hydrodynamics provides a resummation \eqref{eq:HydroCorrelator} that is valid at sufficiently small $\omega$ and $k$, and its structure guarantees that the pole skipping point at $\omega_0=0$, $k_0=0$ survives at any non-zero $\bar{\lambda}$. The locations of other pole skipping points depend on the values of the transport coefficients.

\paragraph{}Anticipating that QFTs with a gravitational dual always have a pole skipping point at $\omega_*=i2\pi T$, we can first try to use the universal hydrodynamic dispersion relation to determine the wavenumber at which the hydrodynamic poles have this frequency. Practically, this is most easily done using equation \eqref{eq:resummedhydrodispersion} where the resummation in $k$ has been performed, and yields
\begin{equation}
\label{eq:kstarfromhydro}
k_*^2=-(2\pi T)^2\left(1+2\alpha_\Delta\left(\frac{\pi\Delta(\Delta-2)\cot\left(\frac{\pi\Delta}{2}\right)}{4(\Delta-1)}-1\right)\bar{\lambda}^2+\ldots\right),
\end{equation}
where $\ldots$ denote higher-order corrections in $\bar{\lambda}$ and
\begin{equation}
\alpha_\Delta=\frac{3 (2 \pi )^{2 (\Delta -1)} \Gamma (2-\Delta ) \Gamma \left(\frac{\Delta }{2}\right)^2}{\Gamma \left(1-\frac{\Delta}{2}\right)^2 \Gamma (\Delta )}.
\end{equation}
However this wavenumber lies outside the radius of convergence $k_{\text{eq}}=\Delta\pi T$ of the high temperature hydrodynamic dispersion relation, since any relevant deformation has $\Delta<2$. A sketch of this is shown in Figure~\ref{fig:PoleSkippingPlot} below. The result that a pole passes through the location $(\omega_*,k_*)$ therefore relies on analytically continuing this dispersion relation outside of its radius of convergence using \eqref{eq:resummedhydrodispersion}. 

\paragraph{}What, then, is the limit in which the resummed dispersion relation \eqref{eq:resummedhydrodispersion} can be trusted if it is not just the hydrodynamic limit $k<k_{\text{eq}}$? As weakly breaking conformal symmetry shifts the poles slightly away from the lightcone $\omega_\pm=\pm k$, the natural guess is that it is valid at leading order in the high temperature, near-lightcone expansion
\begin{equation}
\label{eq:nearlightconeexpansion}
 \omega\mp k\sim\bar{\lambda}^2,\quad\quad\quad\quad\quad\quad\quad\quad \bar{\lambda}^2\ll1.   
\end{equation}
Taking the limit \eqref{eq:nearlightconeexpansion} in the full hydrodynamic two-point function \eqref{eq:HydroCorrelator} amounts to replacing $\varepsilon+P$ and $\kappa$ with their CFT values $\pi cT^2/3$ and $1$ respectively, and replacing $\omega\Omega(\omega,k)\rightarrow\pm k\Omega(\pm k,k^2)$. This gives
\begin{equation}
\label{eq:resummedG}
G(\omega,k)\rightarrow\mp\frac{ck}{24\pi }\frac{(2\pi T)^2+k^2+\ldots}{\omega\mp k\left(1+\Gamma_{\pm}(k)\right)+\ldots}, 
\end{equation}
where
\begin{equation}
\begin{aligned}
    \Gamma_{\pm}(k)=-\frac{1}{2}\left(1-c_s^2\mp ik\Omega( \pm k,k^2)\right).
    \end{aligned}
\end{equation}
This is the high temperature, near-lightcone limit of the universal hydrodynamic correlator, in which $\Gamma_\pm(k)$ should be understood as a series expansion in $k$ -- with the finite radius of convergence $k_{\text{eq}}$ --  whose coefficients can be obtained by determining the expansion of $\Omega$ from equation \eqref{eq:generatingfunctional}.

\paragraph{}To continue the result \eqref{eq:resummedG} to shorter scales prior to the emergence of hydrodynamics (i.e.~to $k>k_{\text{eq}}$) we simply resum the series for $\Gamma_\pm(k)$ by evaluating equation \eqref{eq:generatingfunctional} at $\omega=\pm k$. This yields
\begin{equation}
\begin{aligned}
\label{eq:resummedGamma}
    \Gamma_\pm(k)&\,=-\frac{6\pi\lambda^2(2-\Delta)^2}{(2\pi T)^2+k^2}\left(G^{\text{CFT}}_{\mathcal{O}\mathcal{O}}(\pm k,k)-\frac{\Delta}{(2-\Delta)}G^{\text{CFT}}_{\mathcal{O}\mathcal{O}}(0,0)\right),\\
    &\,=-\bar{\lambda}^2\frac{\Delta(2-\Delta)}{2(1-\Delta)}\frac{\alpha_\Delta}{1+\left(\frac{k}{2\pi T}\right)^2}\left(\frac{\Gamma\left(2-\frac{\Delta}{2}\right)\Gamma\left(\frac{\Delta}{2}\mp\frac{ik}{2\pi T}\right)}{\Gamma\left(1+\frac{\Delta}{2}\right)\Gamma\left(1-\frac{\Delta}{2}\mp\frac{ik}{2\pi T}\right)}-1\right),
\end{aligned}
\end{equation}
where on the second line we used the explicit expression for $G^{\text{CFT}}_{\mathcal{O}\mathcal{O}}(\omega,k)$ in equation \eqref{eq:OOCFT}. We now analytically continue this meromorphic function to all $k$ away from the isolated poles $k=q_n=\mp i\pi T(\Delta+2n)$ ($n=0,1,2,\ldots$). While this analytic continuation is straightforward, the original series expansion proposed in \cite{Davison:2024msq} is far from proven. This is why we are testing \eqref{eq:resummedG} in holographic theories.

\paragraph{}The proposal is that \eqref{eq:resummedG}, with $\Gamma_{\pm}$ given by equation \eqref{eq:resummedGamma}, is the two-point function at leading order in the high temperature, near-lightcone limit \eqref{eq:nearlightconeexpansion}. By construction, expanding this at small $k$ yields the universal hydrodynamic correlator near the lightcone, while expanding at small $\bar{\lambda}$ yields 
\begin{equation}
\begin{aligned}
\label{eq:resummedGExpansion}
    G(\omega,k)\rightarrow&\,\mp\frac{ck}{24\pi }\frac{(2\pi T)^2+k^2}{\omega\mp k}\\
    &\,+ \frac{\pi cT^2k^2}{12}\frac{\bar{\lambda}^2}{(\omega\mp k)^2}\frac{\Delta(2-\Delta)\alpha_\Delta}{(1-\Delta)}\left(\frac{\Gamma\left(2-\frac{\Delta}{2}\right)\Gamma\left(\frac{\Delta}{2}\mp\frac{ik}{2\pi T}\right)}{\Gamma\left(1+\frac{\Delta}{2}\right)\Gamma\left(1-\frac{\Delta}{2}\mp\frac{ik}{2\pi T}\right)}-1\right)+\ldots, 
    \end{aligned}
\end{equation}
which agrees with the near-lightcone limit of the conformal perturbation theory result \eqref{eq:CPTstructure} to quadratic order in $\bar{\lambda}$. Indeed, one can obtain \eqref{eq:resummedG} without using \cite{Davison:2024msq} by noting that the $O(\lb^2)$ conformal perturbation theory correction in~\eqref{eq:resummedGExpansion} diverges faster than the leading term near the lightcone, and naively resumming it to \eqref{eq:resummedG}.

\paragraph{}The resummed high temperature, near-lightcone expansion of the correlator breaks down for $k$ within a distance $\sim\bar{\lambda}^2$ of a $q_n$. There is a clear physical interpretation of this: at these wavenumbers there are additional thermal excitations of the QFT that approach the lightcone. As can be seen from the explicit expression \eqref{eq:OOCFT} for $G^{\text{CFT}}_{\mathcal{O}\mathcal{O}}(\omega,k)$, there are thermal excitations of $\mathcal{O}$ that cross the lightcone at exactly $k=q_n$ when $\bar{\lambda}=0$, and we expect this to happen at $k= q_n+O(\bar{\lambda}^2)$ for $\bar{\lambda}\ll1$. When conformal symmetry is broken, operator mixing (see e.g.~\eqref{eq:explicitWardidentities1} and~\eqref{eq:explicitWardidentities2} in Appendix~\ref{app:comparison}) requires these excitations to also appear as poles of the stress tensor correlator. Our proposal breaks down near $q_n$ because it does not include these additional poles. Note that these poles approach the lightcone for imaginary $k$: we should not think of them physically as propagating along the lightcone, but nevertheless they lead to a breakdown of the high-temperature, near-lightcone expansion in complexified momentum space.

\paragraph{}Notice that since $q_0= k_{\text{eq}}$, our resummed expression for the stress tensor correlator breaks down for $k$ parametrically close to where the derivative expansion of universal hydrodynamics breaks down. This is not a coincidence: as explained above we expect that the breakdown of hydrodynamics is caused by a branch point singularity in the dispersion relation near the lightcone, where the hydrodynamic pole `collides' with a pole representing a thermal excitation of $\mathcal{O}$. This branch point would only be visible at higher order in the high temperature, near-lightcone expansion \eqref{eq:nearlightconeexpansion}: as it arises from operator mixing, the residue of the latter pole in the stress tensor correlator is suppressed by a factor of $\bar{\lambda}^2$. Our proposal says that once we increase $k$ past $k_{\text{eq}}$ there should still be a pole with dispersion relation given by \eqref{eq:resummedhydrodispersion}. We expect further branch point singularities when $k$ approaches each $q_n$, due to collisions near the lightcone with poles representing the successively shorter-lived thermal excitations of $\mathcal{O}$, but with a pole with dispersion relation \eqref{eq:resummedhydrodispersion} surviving for all $k$ away from these isolated points.  Although this sounds exotic and perhaps unlikely, it is very similar to what happens in the nearly-conformal, low temperature limit of large-$N$ SYK chains \cite{Choi:2020tdj} and AdS$_2\times$R$^d$ black holes \cite{Arean:2020eus,Gouteraux:2025kta} (see also \cite{Wu:2021mkk,Jeong:2021zsv,Huh:2021ppg,Liu:2021qmt,Gursoy:2021vpu,CruzRojas:2024igr}).

\paragraph{}Finally, we will now use the resummed two-point function to predict locations of pole skipping points in holographic QFTs of this kind. In general these are expected at $\omega=+i2\pi T$, $0$, $-i2\pi T$, $-i4\pi T,\ldots$, but the near-lightcone limit of the two-point function \eqref{eq:resummedG} will only capture the subset of pole skipping points for which $k=\pm \omega+O(\lb^2)$.  From the numerator of \eqref{eq:resummedG} we see that in fact the only two wavenumbers for which this happens (besides the trivial pole skipping point at $\omega_0=0$, $k_0=0$) are $k^2=-(2\pi T)^2+O(\bar{\lambda}^2)$. From the denominator, we can explicitly obtain the correction. 

\paragraph{}First, there is the pole skipping point in the upper half plane at $\omega_*=+i2\pi T$ and the wavenumber $k_*$ identified earlier in \eqref{eq:kstarfromhydro}. Using the expected relation \eqref{eq:vBintro} between $k_*$ and $v_B$ in holographic theories, this predicts
\begin{equation}
\label{eq:hydrovBprediction}
    v_B=1-\alpha _{\Delta } \left(\frac{\pi \Delta  (\Delta -2)   \cot
   \left(\frac{\pi  \Delta }{2}\right)}{4 (\Delta -1)}-1\right) \bar{\lambda }^2+\ldots ,
\end{equation}
in these theories. For any relevant deformation this is subluminal and supersonic $c_s<v_B<1$ where the speed of sound is $c_s=1-(2-\Delta)\alpha_\Delta\bar{\lambda}^2+\ldots$ \cite{Delacretaz:2021ufg}. Second, there is a pole skipping point in the lower half plane at $\omega=-i2\pi T$ and the wavenumber $k_1$ where
\begin{equation}
\label{eq:HydroLHPpoleskippingk}
    k_1^2=-(2\pi T)^2\left(1-2\alpha_\Delta\left(\frac{\pi\Delta(\Delta-2)\cot\left(\frac{\pi\Delta}{2}\right)}{4(\Delta-1)}+1\right)\bar{\lambda}^2\right)+\ldots.
\end{equation}
See Figure~\ref{fig:PoleSkippingPlot} below for a sketch of this. These are specific and non-trivial predictions for holographic theories, and in the remainder of the paper we will verify that they are true. Note that we assumed, rather than derived, the locations of the pole skipping frequencies based on holographic expectations. Essentially, after doing this we used the denominator of \eqref{eq:resummedG} to obtain the wavenumber at which a pole passes through this frequency, and used the numerator as a non-trivial check that -- to the order to which \eqref{eq:resummedG} is valid -- it is consistent that a zero also passes through this point. To go beyond this and truly prove that a zero passes through this point would require knowing the first correction to the numerator of \eqref{eq:resummedG} (at least in the vicinity of the pole skipping points). Equivalently, knowing this correction would allow us to derive -- rather than assume -- the locations of the pole skipping frequencies. Obtaining these corrections from a resummation of universal hydrodynamics is beyond the scope of this work.

\begin{figure}[h]
    \centering
    \includegraphics[width=.6\textwidth]{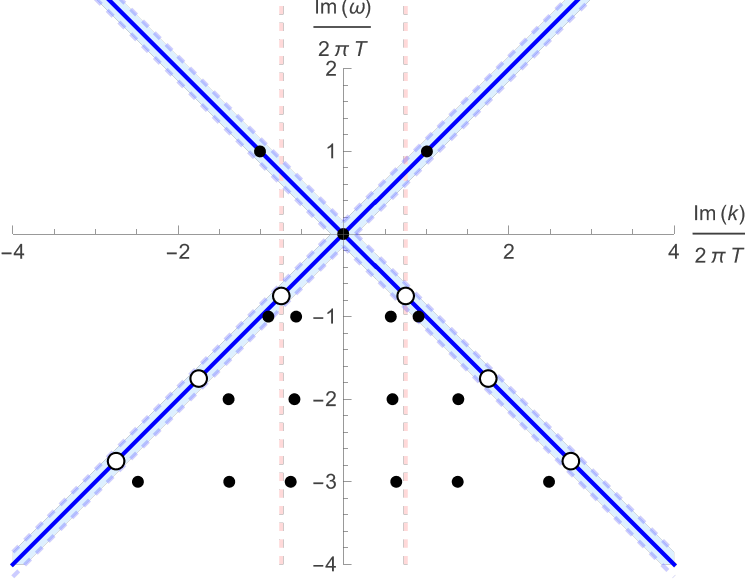}
    \qquad
    \caption{    
    A sketch of the pole skipping points (black dots) on the imaginary $(\omega,k)$ plane for $\bar{\lambda}\ll1$ and $\Delta=3/2$. The near-lightcone region is shaded in light blue, the red lines show the radius of convergence $k_{\text{eq}}$ of the hydrodynamic dispersion relation and the white circles show the wavenumbers $q_n$ where the high temperature near-lightcone expansion breaks down. The only pole-skipping points near the lightcone are those at $\omega=0,
    \pm i2\pi T$. Those at $\omega=\pm i2\pi T$ lie outside the radius of convergence of the hydrodynamic dispersion relation but can be captured by an analytic continuation. The remaining pole skipping points are inherited from $G_{\mathcal{O}\mathcal{O}}^{\text{CFT}}$ due to operator mixing.\label{fig:PoleSkippingPlot}}
\end{figure}

\paragraph{}The near-lightcone pole skipping wavenumbers we have just identified agree with those of the CFT stress tensor two point function, with small $\bar{\lambda}^2$ corrections. The $\mathcal{O}$ two-point function in a CFT \eqref{eq:OOCFT} also exhibits pole skipping at infinitely many frequencies in the lower half-plane \cite{Blake:2019otz,Grozdanov:2019uhi}
\begin{equation}
\label{eq:CFTscalarPSpts}
\omega_n=-i2\pi T,\quad\quad\quad\quad\text{and}\quad\quad\quad\quad k_{n,q}=\pm i2\pi T(n-2q+\Delta),
\end{equation}
where $n=1,2,3,\ldots$ and $q=1,2,\ldots,n$. Due to operator mixing, we expect the stress tensor two-point function to inherit a similar family of pole skipping points when conformal symmetry is weakly broken $\bar{\lambda}\ll1$. However, since it is not the hydrodynamic mode that passes through them, we cannot access them from hydrodynamics or its near-lightcone resummation. We will derive expressions for the first few such pole skipping wavenumbers directly in holographic theories later, and see that they are sensitive to OPE coefficients of the operator $\mathcal{O}$ as well as its dimension. A sketch of the locations of these pole skipping points is shown in Figure~\ref{fig:PoleSkippingPlot}.

\section{Butterfly velocity of holographic theories}
\label{sec:Equilibriumsolution}

\paragraph{}We are now going to consider (1+1)-dimensional holographic QFTs. We will determine the spacetime corresponding to the thermal equilibrium state to quadratic order in the high temperature expansion $\bar{\lambda}\ll1$ and then use this to compute the butterfly velocity in the same limit. 

\subsection{The thermal state}

\paragraph{}The simplest holographic versions of high temperature (1+1)d QFTs are captured by the action
\begin{equation}
\label{eq:bulkaction}
    S=\frac{1}{16\pi G}\int d^3x\sqrt{-g}\left(R-\frac{1}{2}\partial_M\phi\partial^M\phi+V(\phi)\right)+S_{\text{boundary}},
\end{equation}
where the boundary terms $S_{\text{boundary}}$ are discussed in \cite{deHaro:2000vlm} and the scalar potential is chosen such that at small $\phi$
\begin{equation}
\label{eq:scalarPotential}
    V(\phi\rightarrow0)=2-\frac{1}{2}\Delta(\Delta-2)\phi^2+V_3\phi^3+V_4\phi^4+\ldots,
\end{equation}
where $0<\Delta<2$. When $\phi=0$ this action has AdS$_3$ as a classical solution, corresponding to a CFT with central charge $c=3/(2G)$. We have set the AdS radius to $L=1$.

\paragraph{}We parameterise the thermal states of these QFTs by planar black hole solutions of the form
\begin{equation}
\label{eq:equilibriummetric}
ds^2=-D(r)dt^2+C(r)dx^2+B(r)dr^2,\quad\quad\quad\quad\quad\quad\quad\phi=\Phi(r).
\end{equation}
We assume that $\Phi'(r)\ne0$, which corresponds to the breaking of conformal symmetry. The classical equations of motion for this ansatz can be written
\begin{equation}
    \label{eq:BGEoMs}
    \left(\log\left(\frac{C'}{\sqrt{BCD}}\right)\right)'=-\frac{C{\Phi'}^2}{C'},\quad \left(\frac{C^{3/2}(D/C)'}{\sqrt{BD}}\right)'=0, \quad    \frac{C'D'}{CD}={\Phi'}^2+2BV(\Phi),
\end{equation}
where primes denote derivatives with respect to $r$. Notice that there are only three equations for the four fields $B$, $C$, $D$ and $\Phi$. We will shortly choose a useful radial coordinate to remove this redundancy. The above equations of motion can be combined to obtain the following relations between the fields
\begin{equation}
\begin{aligned}
\label{eq:BGEoMs2}
    \left(\sqrt{\frac{CD}{B}}\Phi'\right)'&\,=-\sqrt{BCD}\left.\frac{dV}{d\phi}\right|_{\phi=\Phi},\quad\quad\quad \left(\sqrt{\frac{D}{BC}}C'\right)'=2\sqrt{BCD}V(\Phi).
\end{aligned}    
\end{equation}
These relations will be useful soon.

\paragraph{}We are interested in solutions with a planar horizon at $r=r_0>0$, near which the solution can be written
\begin{equation}
\begin{aligned}
\label{eq:BGhorBCs}
    B(r\rightarrow r_0)&\,\rightarrow \frac{b}{4\pi T(r_0-r)}+\ldots,\quad\quad\quad\quad C(r\rightarrow r_0)\rightarrow (4Gs)^2+\ldots,\\
    D(r\rightarrow r_0)&\,\rightarrow 4\pi Tb(r_0-r)+\ldots,\quad\quad\quad\quad \Phi(r)\rightarrow\Phi_0+\ldots,
    \end{aligned}
\end{equation}
where $b>0$, $\Phi_0$, $s$ and $T>0$ are constants. From the Bekenstein-Hawking formula, and regularity of the Euclidean solution, the latter two constants are the entropy density and temperature of the thermal state. We also demand that the solutions are asymptotically AdS$_3$ and correspond to CFTs deformed by a scalar operator of dimension $\Delta$ as in \eqref{eq:deformedaction}. In practice this means that as one approaches the asymptotic boundary $r\rightarrow0$ the metric functions are
\begin{equation}
\begin{aligned}
\label{eq:BGbdyBCs}
    B(r\rightarrow0)\rightarrow\frac{1}{r^2}+\ldots,\quad\quad\quad C(r\rightarrow0)\rightarrow\frac{1}{r^2}+\ldots,\quad\quad\quad D(r\rightarrow0)\rightarrow\frac{1}{r^2}+\ldots,
    \end{aligned}
\end{equation}
while the asymptotics of the scalar field depend upon the value of $\Delta$. For $1<\Delta<2$, the leading term is
\begin{equation}
\label{eq:PhiAdS3asymptotics}
    \Phi(r\rightarrow0)\rightarrow\frac{\sqrt{12}\pi}{1-\Delta}\lambda r^{2-\Delta}+\ldots,
\end{equation}
where $\lambda$ is the constant source for $\mathcal{O}$. For $0<\Delta<1$, the leading term in the near-boundary expansion is $\sim r^\Delta$, but the appropriate boundary condition is still \eqref{eq:PhiAdS3asymptotics}. For $\Delta=1$ the leading term has a logarithmic divergence \cite{Klebanov:1999tb} and we will not consider this case explicitly.

\paragraph{}To find the black hole solutions dual to thermal states of these deformed CFTs, the equations of motion \eqref{eq:BGEoMs} should be solved subject to the above boundary conditions. This results in relations between the constants in the near-horizon and near-boundary solutions, such as $s(T,\lambda)$. In general the equations can only be solved numerically (see \cite{Ecker:2020gnw,Jiang:2024tdj,Frenkel:2020ysx,Caceres:2021fuw} for studies of such numerical solutions).

\subsection{High temperature solution and thermodynamics}

\paragraph{}Although exact solutions for the thermal state can only be obtained numerically, we can make progress analytically by working perturbatively in the high temperature limit $\bar{\lambda}\ll1$. Since we are concerned with the equilibrium solution, we do not need to worry here about the late-time breakdown of high temperature perturbation theory. When $\lambda\ne0$, the scalar field $\Phi(r)$ has a non-trivial profile that backreacts on the BTZ metric and diverges logarithmically in the interior~\cite{Caceres:2021fuw}. In the high temperature limit $\bar{\lambda}\ll1$, the horizon cloaks this region such that $\Phi$ is small everywhere outside it. 

\paragraph{}We first fix the gauge by choosing the radial coordinate such that
\begin{equation}
\label{eq:Cgaugechoice}
    C(r)=\frac{1}{r^2},
\end{equation}
and then expand the remaining fields $B$, $D$ and $\Phi$ as series in $\bar{\lambda}$. At leading order in this expansion the metric functions are those of the BTZ black brane
\begin{equation}
    B_{\text{BTZ}}(r)=\frac{1}{r^2f(r)},\quad\quad\quad D_{\text{BTZ}}(r)=\frac{f(r)}{r^2},\quad\quad\quad f(r)=1-\frac{r^2}{r_0^2},
\end{equation}
while $\Phi$ satisfies the first equation in \eqref{eq:BGEoMs2} in the BTZ metric. The solution for $\Phi$ obeying the boundary conditions described above is
\begin{equation}
\label{eq:Phiperturbativesol}
    \Phi(r)=\Phi_0\,{}_2F_{1}\left(1-\frac{\Delta}{2},\frac{\Delta}{2},1;1-\frac{r_0^2}{r^2}\right),
\end{equation}
where
\begin{equation}
\label{eq:Phi0LambdaRelation}
        \Phi_0=-\sqrt{12}\pi\frac{\Gamma(\frac{\Delta}{2})^2}{\Gamma(\Delta)}\lambda r_0^{2-\Delta}.
\end{equation}
As the temperature of the BTZ solution is $T=1/(2\pi r_0)$, we see that the $\bar{\lambda}$ expansion is just an expansion in the amplitude of the scalar field $\Phi_0$.

\paragraph{}The scalar field \eqref{eq:Phiperturbativesol} backreacts and gives corrections to the metric at $O(\bar{\lambda}^2)$. We now compute these explicitly in a series of steps. First take the first equation in \eqref{eq:BGEoMs} and integrate to obtain
\begin{equation}
\begin{aligned}
    \sqrt{BD}&\,=\frac{1}{r^2}\left(1+c_1-\frac{1}{2}\int_0^r dr r{\Phi'}^2\right),
    \end{aligned}
\end{equation}
at the required order, where $c_1$ is an integration constant. The integrand is regular at the horizon. This is not enough to determine the corrections to $B$ and $D$ individually. To obtain those we now take the second equation in \eqref{eq:BGEoMs2} which, in our gauge and to quadratic order in $\Phi$, can be written
\begin{equation}
    \frac{d}{dr}\left(\frac{D}{r^2\sqrt{BD}}\right)=-\frac{2}{r}\sqrt{BD}+\frac{m^2}{2r^2}\sqrt{BD}\Phi^2.
\end{equation}
This can be integrated to give an expression for $D$ in terms of integrals of $\Phi$. After a non-trivial integration by parts, and using properties of the solution \eqref{eq:Phiperturbativesol} for $\Phi$, this gives
\begin{equation}
    D=\frac{f(r)}{r^2}+\frac{1}{r^2}\left[2c_1+\frac{1}{2}c_2 r^2+\frac{r^2}{2r_0^2}\int_0^rdr r{\Phi'}^2-\frac{1}{2r_0^2}\int^r_0 dr r^3{\Phi'}^2\right],    
\end{equation}
where $c_2$ is an integration constant.

\paragraph{}We have now determined both metric functions $B$ and $D$ individually and are ready to impose boundary conditions to fix the integration constants. Keeping the boundary field theory metric fixed requires $c_1=0$. To fix $c_2$, we choose the location of the horizon to remain at $r=r_0$, even after backreacting the scalar. This fixes
\begin{equation}
        c_2=\frac{1}{r_0^2}\int^{r_0}_0\left(\frac{r^2}{r_0^2}-1\right)r{\Phi'}^2dr.
\end{equation}
Substituting in the explicit solution \eqref{eq:Phiperturbativesol} for $\Phi$ and changing variables to $z=\frac{r_0^2}{r^2}-1$ gives
\begin{equation}
    \begin{aligned}
    c_2&\,=-\frac{\Phi_0^2\Delta^2\left(1-\frac{\Delta}{2}\right)^2}{2r_0^2}\int^\infty_0dz\, {}_2F_{1}\left(2-\frac{\Delta}{2},1+\frac{\Delta}{2},2;-z\right)^2zdz\\
    &\,=\Phi_0^2\frac{(\Delta-1)\tan\left(\frac{\pi\Delta}{2}\right)}{\pi r_0^2},
    \end{aligned}
\end{equation}
where the definite integral was evaluated using \texttt{Mathematica 12}.

\paragraph{}We have now computed the equilibrium state to quadratic order in $\bar{\lambda}$. As a consistency check, we extract from this the entropy density $s$, for which the corresponding result can also be computed directly from the field theory action \eqref{eq:deformedaction} using traditional Euclidean conformal perturbation theory on the cylinder (see e.g.~\cite{Delacretaz:2021ufg}). To do this, we first obtain the entropy density as a function of the horizon radius from the Bekenstein-Hawking formula
\begin{equation}
\label{eq:4Gs}
    s=\frac{1}{4G}\frac{1}{r_0}=\frac{c}{6r_0}.
\end{equation}
This expression is fixed by our choice of gauge \eqref{eq:Cgaugechoice} and so is not on its own physical. The physical information comes from computing the temperature $T(r_0,\bar{\lambda}^2)$ in this gauge as
\begin{equation}
        4\pi T=-\left(\frac{d}{dr}\sqrt{\frac{D}{B}}\right)_{r=r_0}=\frac{2}{r_0}\left(1-\Phi_0^2\frac{(\Delta-1)\tan\left(\frac{\pi\Delta}{2}\right)}{2\pi}\right),
\end{equation}
and then inverting this to obtain
\begin{equation}
\label{eq:r0correction}
    r_0=\frac{1}{2\pi T}\left(1+\alpha_\Delta\bar{\lambda}^2\right),
\end{equation}
to the order we are working at. Substituting this into the Bekenstein-Hawking formula yields
\begin{equation}
\label{eq:entropyCPT}
    s(T,\bar{\lambda}^2)=\frac{\pi cT}{3}(1-\alpha_\Delta\bar{\lambda}^2+\ldots).
\end{equation}
This agrees with the result obtained from Euclidean conformal perturbation theory (see e.g.~equation (16) of \cite{Delacretaz:2021ufg}).

\subsection{The butterfly velocity}

\paragraph{}Using this equilibrium solution we can now determine the butterfly velocity $v_B$ characterising the spread of out-of-time-ordered correlations. In holographic theories $v_B$ is related to the scattering of two particles near the black hole horizon, and can be calculated from the form of the shockwave geometry sourced by one such particle \cite{Shenker:2013pqa,Shenker:2014cwa,Roberts:2014isa,Maldacena:2015waa,Dray:1984ha,Sfetsos:1994xa}. Ultimately this is controlled by the near-horizon metric, with a general expression for $v_B$ given in \cite{Blake:2016wvh,Roberts:2016wdl}. In coordinates where the equilibrium metric is of the form \eqref{eq:BGhorBCs} near the horizon
\begin{equation}
\label{eq:vBmetric}
v_B^2=\lim_{r\rightarrow r_0}\left(\frac{D'}{C'}\right).
\end{equation}

\paragraph{}Before evaluating this explicitly for the high temperature equilibrium solution, we first perform a sanity check by proving that $v_B$ is subluminal for any sensible black hole solution. With the asymptotically AdS boundary conditions \eqref{eq:BGbdyBCs}, $D'/C'\rightarrow1$ near the boundary $r\rightarrow0$. In general, this quantity runs as we move inwards from the boundary to the horizon and thus $v_B^2$ defined in \eqref{eq:vBmetric} is not $1$. The running is controlled by the equation
\begin{equation}
\label{eq:runningBGeq}
    \left(\frac{D'}{C'}\right)' =\left(\frac{C{\Phi'}}{{C'}}\right)^2\left(\frac{D}{C}\right)',
\end{equation}
which follows from the classical equations of motion for the equilibrium solution \eqref{eq:BGEoMs}. Furthermore, the second equation in \eqref{eq:BGEoMs} is a radial conservation equation which when evaluated on the horizon \eqref{eq:BGhorBCs} gives
\begin{equation}
    \left(\frac{D}{C}\right)'=-16\pi G sT\frac{\sqrt{BD}}{C^{3/2}}.
\end{equation}
Assuming that $B$, $C$ and $D$ are all positive outside the horizon then we deduce from equation \eqref{eq:runningBGeq} that $\left(D'/C'\right)'\leq 0$ everywhere outside the horizon and therefore the running is such that $v_B^2\leq 1$ with equality in the conformal limit $\Phi'=0$. See \cite{Mezei:2016zxg} for related bounds on $v_B$ in general dimensions from the null energy condition.

\paragraph{}We can similarly check that any sensible black hole solutions will also have $v_B^2\geq 0$. From the expression \eqref{eq:vBmetric}, and the near-horizon form of the metric \eqref{eq:BGhorBCs}, this requires $C'(r_0)<0$. From integrating the first field equation in \eqref{eq:BGEoMs} and imposing the asymptotically AdS$_3$ conditions in \eqref{eq:BGbdyBCs} we find that
\begin{equation}
    C'=-2\sqrt{BCD}\exp\left(-\int^{r}_0 \frac{C{\Phi'}^2}{C'}dr\right).
\end{equation}
Since the exponent is real, $C'(r_0)$ will be negative for the near-horizon solutions in \eqref{eq:BGhorBCs}.

\paragraph{}We now explicitly evaluate $v_B^2$ to quadratic order in $\bar{\lambda}$, using the equilibrium solution constructed above. Substituting into the expression \eqref{eq:vBmetric} and keeping terms to the relevant order gives
\begin{equation}
\begin{aligned}
\label{eq:holographicvB}
v_B^2&\,=1-\Phi_0^2\left(1-\frac{\Delta}{2}\right)^2\frac{\Delta^2}{4}\int_{0}^\infty dz\,{}_2F_{1}\left(2-\frac{\Delta}{2},1+\frac{\Delta}{2},2;-z\right)^2\\
&\,=1-2 \alpha_\Delta\left(\frac{\pi\Delta(\Delta-2)\cot\left(\frac{\pi\Delta}{2}\right)}{4(\Delta-1)}-1\right)\bar{\lambda}^2+\ldots,
\end{aligned}
\end{equation}
where the integral was evaluated using \texttt{Mathematica 12}. This expression agrees exactly with that found in \eqref{eq:hydrovBprediction} above from the near-lightcone resummation of the universal hydrodynamics of \cite{Davison:2024msq}. It agrees with the holographic calculation of \cite{Jiang:2024tdj} when $\Delta=3/2$ and the numerical holographic results for general $\Delta$ in \cite{Asplund:2025nkw}.

Recall that to obtain this result from hydrodynamics we assumed that we could read off $v_B$ from the location of a pole skipping point -- we will verify this for (1+1)d holographic QFTs shortly.

\section{Pole skipping in holographic theories}
\label{sec:PoleSkipping}

\paragraph{}In this Section we are going to identify the locations of pole skipping points in the thermal two-point function of the stress tensor in the equilibrium states of theories with the gravitational action \eqref{eq:bulkaction}. These are the set of isolated points in momentum space $(\omega,k)$ at which the dispersion relation of a pole $\omega_p(k)$ and of a zero $\omega_z(k)$ of the retarded two-point function intersect. We will first confirm that -- as expected -- there is a pole skipping point located in the upper half frequency plane at
\begin{equation}
\label{eq:UHPpspointsec4}
    \omega_*=i2\pi T,\quad\quad\quad\quad\quad\quad k_*^2=-\left(\frac{2\pi T}{v_B}\right)^2,
\end{equation}
where $v_B$ is given by equation \eqref{eq:vBmetric}. Combined with the small $\bar{\lambda}$ expression obtained for $v_B^2$ in \eqref{eq:holographicvB}, this confirms that the pole skipping point in the upper half-plane agrees with that predicted by the near-lightcone resummation of universal hydrodynamics. We will then investigate the wavenumbers of the pole skipping points in the lower half frequency plane.\footnote{As the basic structure of large $c$ hydrodynamics guarantees there will be a pole skipping point at $\omega=k=0$ (see equation \eqref{eq:HydroCorrelator}) we will not attempt to re-derive this holographically.} For small $\bar{\lambda}$ we confirm that one of these (at $\omega=-i2\pi T$) agrees with the other predicted pole skipping point of the near-lightcone resummation of universal hydrodynamics \eqref{eq:HydroLHPpoleskippingk}. We find explicit expressions for the next few pole skipping points closest to $\omega=0$. These are qualitatively different to those accessible from resummed hydrodynamics: they are far from the lightcone at high temperatures, and their wavenumbers are non-universal in that they depend on OPE coefficients of the operator $\mathcal{O}$ as well as its dimension.

\subsection{Pole skipping in upper half-plane}
\label{sec:UHPpoleskippingholo}

\paragraph{}In the gravitational description of a QFT, the origin of pole skipping at isolated points $(\omega,k)$ in momentum space is the existence of an additional ingoing solution to the equations of motion for small amplitude perturbations of bulk fields with these $(\omega,k)$. This extra solution means that the correlator is undefined at this point in momentum space. More precisely, it means the correlator is infinitely multi-valued around this point and so it can be understood as the location where a line of poles in the correlator intersects with a line of zeroes \cite{Blake:2018leo,Blake:2019otz}.

\paragraph{}The locations of pole skipping points are aspects of the non-equilibrium response that can be determined without having to solve for the propagation of fields from the spacetime's boundary to its horizon. For this reason they are theoretically attractive, requiring only local solutions of the equations of motion near the horizon. Furthermore, the locations are directly related to the equilibrium spacetime near the horizon and so can easily be evaluated in a high temperature expansion $\bar{\lambda}\ll1$ using the results of Section \ref{sec:Equilibriumsolution}.

\paragraph{}For the case of pole skipping in stress tensor thermal two-point functions, the relevant equations are the Einstein equations for small perturbations of the metric. After transforming from the coordinate $t$ to the ingoing coordinate
\begin{equation}
\label{eq:vcoorddefn}
    v=t-\int_0^r dr' \sqrt{\frac{B(r')}{D(r')}},
\end{equation}
we denote the small amplitude perturbations of the metric as $\delta g_{MN}$.

\paragraph{}In the conformal limit, the general solutions to these equations are simply linearised gauge transformations. When conformal symmetry is broken ($\Phi'(r)\ne0$), perturbations of the metric couple to those of the scalar field $\delta\phi$ such that the general solution now also contains a gauge-invariant propagating degree of freedom. The complete set of equations for the perturbations is given in Appendix \ref{app:perturbationequations}.

\paragraph{}Working in radial gauge $\delta g_{Mr}=0$, we call an ingoing solution one for which both $\delta g_{\mu\nu}$ and $\delta\phi$ have Taylor series expansions in $(r_0-r)$. These solutions can be constructed locally by similarly expanding the equations of motion and then solving order-by-order. A generic such solution is parameterised by four independent functions of $(v,x)$. These encode the perturbations of the three components of the QFT metric plus the perturbation of the source for the scalar operator $\mathcal{O}$.

\paragraph{}The existence of an additional ingoing solution can be identified by examining the equations of motion in the vicinity of the horizon. In higher-dimensional cases, the generic pole skipping point in the upper half plane at \eqref{eq:UHPpspointsec4} is most easily seen by directly examining the $vv$ component of the Einstein equation on the horizon \cite{Blake:2018leo}. The analysis of \cite{Blake:2018leo} follows through in a straightforward manner for the (2+1)-dimensional gravitational theory \eqref{eq:bulkaction} as we now briefly summarise. In radial gauge, the $vv$ component of the Einstein equation evaluated on the horizon is
\begin{equation}
    \begin{aligned}
    \label{eq:Evvhorizon}
\left(\omega-i2\pi T\right)\left(\omega\delta g_{xx}(r_0)+2k\delta g_{vx}(r_0)\right)+\left(k^2-i\omega\frac{2\pi T}{v_B^2}\right)\delta g_{vv}(r_0)=0,
    \end{aligned}
\end{equation}
where $v_B$ is given by equation \eqref{eq:vBmetric} and we have assumed that the solution is ingoing. For a generic $(\omega,k)$, equation \eqref{eq:Evvhorizon} provides a condition relating the values of $\delta g_{vv}(r_0)$ and $\omega\delta g_{xx}(r_0)+2k\delta g_{vx}(r_0)$ in any ingoing solution. However, at the special point $(\omega_*,k_*)$ this Einstein equation vanishes identically: there is one less condition to satisfy, and therefore one extra ingoing solution, at this point.

\subsection{Master field formalism}

\paragraph{}In principle, the infinitely many pole skipping points expected in a holographic theory can be determined by carrying out the analysis above systematically. At each successively higher order in the near-horizon expansion of the equations of motion, one can identify successively lower frequencies in the complex plane for which one of the equations is identically satisfied for specific choices of wavenumbers \cite{Blake:2019otz}. However, implementing this in practice when the Einstein equations are involved can be cumbersome simply because there are so many of them (see \cite{Wang:2022mcq,Ahn:2024gjh,Loganayagam:2022teq} for other approaches to this).

\paragraph{}To streamline this calculation we will work with a carefully chosen `master field', rather than with the fundamental perturbations of the metric and scalar field themselves. We define the master field as follows
\begin{equation}
\label{eq:Hdefn}
    H(r)=\frac{1}{C(r)}\left(\omega^2 \delta g_{xx}+2 \omega k \delta g_{vx}+k^2\delta g_{vv}-\frac{C'}{\Phi '}\left(\omega ^2-\frac{D'}{C'}k^2\right)\delta \phi\right).
\end{equation}
The key property of $H$ is that it is invariant under the linearised gauge transformations
\begin{align}
\label{eq:residualGTs}
    \delta g_{MN} \rightarrow \delta g_{MN}+\nabla_{M} \xi_{N}+\nabla_{N} \xi_{M},\quad\quad\quad\quad\quad\quad \delta \phi \rightarrow \delta \phi+\xi^{M} \partial_M \Phi.
\end{align}
Of the four independent functions of $(v,x)$ that parameterise the general ingoing solution to the equations of motion in radial gauge, three are simply the residual gauge transformations, for which the exact solutions are
\begin{equation}
    \begin{aligned}
        \delta g_{vv}=&\,-\left(\frac{D'}{\sqrt{BD}}+2
        \partial_v\right)\left(\xi_v-2\left(\frac{\sqrt{BD}C}{C'}-\int dr\frac{\sqrt{BD}C^2{\Phi'}^2}{{C'}^2}\right)\partial_v\xi_r\right)\\
        &\,-2D\partial_v\xi_r,\\
        \delta g_{vx}=&\,2\left(C\int\frac{\sqrt{BD}C{\Phi'}^2}{{C'}^2}dr-\int\frac{\sqrt{BD}C^2{\Phi'}^2}{{C'}^2}dr\right)\partial_x\partial_v\xi_r-\partial_x\xi_v-C\partial_v\xi_x\\
        &\,-D\partial_x\xi_r,\\
        \delta g_{xx}=&\,\frac{C'}{\sqrt{BD}}\xi_v-2C\partial_x\xi_x-2\left(C-\frac{C'}{\sqrt{BD}}\int\frac{\sqrt{BD}C^2{\Phi'}^2}{{C'}^2}dr\right)\partial_v\xi_r\\
        &\,-4\left(\frac{\sqrt{BD}C}{C'}-C\int\frac{\sqrt{BD}C{\Phi'}^2}{{C'}^2}dr\right)\partial_x^2\xi_r,\\
        \delta \phi=&\,\frac{\Phi'}{\sqrt{BD}}\left(\xi_v-2\left(\frac{\sqrt{BD}C}{C'}-\int\frac{\sqrt{BD}C^2{\Phi'}^2}{{C'}^2}dr\right)\partial_v\xi_r\right),
    \end{aligned}
\end{equation}
for any $\xi_M(v,x)$. In this sense, there is only generically one non-trivial ingoing solution to the equations of motion and this is what $H$ captures. To explicitly obtain the solutions for the fundamental metric perturbations, one should first solve equation \eqref{eq:Hdefn} for $\delta\phi$ and then substitute this into the three Einstein equations in radial gauge that are first order in radial derivatives \eqref{eq:EinsteinEqsFirstOrder}. These should then be solved for $\delta g_{\mu\nu}$ (where we use the Greek indices $\mu,\nu,\ldots$ to denote the field theory coordinates $v$ and $x$).

\paragraph{}To determine the non-trivial solution, we must solve $H$'s equation of motion. After using the equilibrium equations of motion, in momentum space this equation may be written 
\begin{equation}
    \begin{aligned}
        \label{eq:H-EOM}
        0=&\,\left(\frac{d}{dr}+i\omega\sqrt{\frac{B}{D}}\right)\left(\frac{\frac{D}{B}\frac{C{\Phi'}^2}{C'}}{\left(\omega^2-k^2\frac{D'}{C'}\right)}\left(\frac{d}{dr}+i\omega\sqrt{\frac{B}{D}}\right)\frac{H}{\sqrt{D/C}}\right)\\
        &\,+\frac{\left(\omega^2-k^2\frac{D}{C}\right)}{\left(\omega^2-k^2\frac{D'}{C'}\right)}\frac{C{\Phi'}^2}{C'}\frac{\frac{D}{B}\frac{C{\Phi'}^2}{C'}}{\left(\omega^2-k^2\frac{D'}{C'}\right)}\left(\frac{d}{dr}+i\omega\sqrt{\frac{B}{D}}\right)\frac{H}{\sqrt{D/C}}\\
        &\,+\frac{B}{D}\left(\omega^2-k^2\frac{D}{C}+\frac{1}{C}\left(\frac{C^{3/2}(D/C)'}{2\sqrt{BD}}\right)^2\right)\frac{\frac{D}{B}\frac{C{\Phi'}^2}{C'}}{\left(\omega^2-k^2\frac{D'}{C'}\right)}\frac{H}{\sqrt{D/C}}.
    \end{aligned}
\end{equation}
This is a second order equation, and at the horizon there is generically one solution that is ingoing and one that is outgoing. It is the contribution of $H$ to $\delta g_{\mu\nu}$ that is responsible for dissipation in the QFT: in the conformal limit $\Phi'(r)=0$ it is clear from \eqref{eq:Hdefn} that $H$ decouples from $\delta g_{\mu\nu}$, whose general solutions are then just given by the residual gauge transformations in \eqref{eq:residualGTs}. Analogues of $H$ have been used for many years to simplify the study of stress tensor dynamics in higher-dimensional holographic theories \cite{Kovtun:2005ev}. The subtlety in (2+1) bulk dimensions is that its existence requires the equilibrium solution to have a matter field that breaks the conformal symmetry.

\paragraph{}Having introduced $H$, we are first going to revisit the pole skipping point at $\omega_*=+i2\pi T$ and check that we can also identify this from the equation of motion for $H$. At a generic $(\omega,k)$, the two independent solutions of \eqref{eq:H-EOM} near the horizon are of the form
\begin{equation}
\label{eq:HgenericBCs}
    H_{\text{in}}\sim (r_0-r)^0,\quad\quad\quad\quad\quad\quad H_{\text{out}}\sim(r_0-r)^{\frac{i \omega}{2\pi T}}.
\end{equation}
The first is ingoing at the horizon and the second is outgoing. At the expected pole skipping frequency $\omega_*=+i2\pi T$, $H_{\text{out}}$ naively diverges near the horizon. However this generic result does not hold when the wavenumber $k_*$ is simultaneously tuned to the expected pole skipping value \eqref{eq:UHPpspointsec4}. In this case the factor $\omega^2-k^2\frac{D'}{C'}$ present in the equation of motion of $H$ vanishes on the horizon, changing the structure of the solutions there. After first setting $\omega=\omega_*$ and $k=k_*$, one finds that the general solution for $H$ near the horizon is actually
\begin{equation}
\label{eq:HsolUHP}
    H=e^{-i\omega_*v+ik_*x}\left(H_0+H_1(r_0-r)+H_2(r_0-r)^2+\ldots\right), 
\end{equation}
where the higher order terms in the near-horizon expansion are fixed uniquely in terms of the arbitrary $H_0$ and $H_1$. In other words, at $(\omega_*,k_*)$ we find that both solutions for $H$ -- rather than just one -- have a Taylor series expansion in ingoing coordinates at the horizon.

\paragraph{}This is the intuitive demonstration that there is an extra ingoing solution in $H$ for modes with $\left(\omega_*,k_*\right)$. To truly confirm this, we have to check that regularity of these modes of the master field $H$ at the horizon indeed corresponds to the fundamental perturbations $\delta g_{\mu\nu}$ and $\delta\phi$ with frequency $\omega_*$ and wavenumber $k_*$ being ingoing there. To verify this explicitly, we first construct the general ingoing solution for fundamental perturbations in radial gauge at $\omega=\omega_*$ and $k=k_*$ by solving the fundamental equations of motion order-by-order near the horizon for the Taylor series coefficients of these perturbations. This general ingoing solution is then substituted into the definition of $H$ \eqref{eq:Hdefn} where it corresponds to the solution \eqref{eq:HsolUHP} with
\begin{equation}
\begin{aligned}
\label{eq:H0H1explicit}
    H_0&\,=\frac{i2\pi T}{(4Gs)^2}\left(2k_*\delta g_{vx}(r_0)+\omega_*\delta g_{xx}(r_0)+i\frac{2\pi T}{v_B^2}\delta g_{vv}(r_0) \right),\\
    H_1&\,=\frac{b}{2}\left(-\frac{1}{(4Gs)^2}\frac{2\pi T}{v_B^2}H_0+\frac{\Vp^2 }{16\pi T}\delta g_{vv}(r_0)\right).
\end{aligned}
\end{equation}
This shows that indeed the two Taylor series solutions for $H$ at $(\omega_*,k_*)$ correspond to two different ingoing solutions for the fundamental perturbations $\delta g_{\mu\nu}$ and $\delta\phi$. Furthermore, the specific expressions for $H_0$ and $H_1$ in \eqref{eq:H0H1explicit} are instructive. Recall from the Einstein equation \eqref{eq:Evvhorizon} that, at a generic point in momentum space, the combinations $2k\delta g_{vx}(r_0)+\omega\delta g_{xx}(r_0)$ and $\delta g_{vv}(r_0)$ are not linearly independent. Therefore we see that the fact that this equation vanishes identically at $(\omega_*,k_*)$ -- such that $2k_*\delta g_{vx}(r_0)+\omega_*\delta g_{xx}(r_0)$ and $\delta g_{vv}(r_0)$ are linearly independent -- is crucial. Otherwise $H_0$ and $H_1$ in \eqref{eq:H0H1explicit} would not be linearly independent and there would be no extra ingoing solution for $H$ at this frequency and wavenumber.

\paragraph{}We have now shown how to identify the pole skipping point with frequency in the upper half-plane by studying the dynamics of the master field $H$. For this particular pole skipping point, it is more complicated than just looking at the Einstein equations for the fundamental fields directly as we did before. The real advantage of the master field formalism is that it simplifies the identification of the pole skipping points with frequency in the lower half-plane, which is what we now turn to.

\paragraph{}Since the basic structure of large $c$ hydrodynamics guarantees there to be pole skipping at $\omega=0$ and $k=0$ we will not investigate this case holographically. This case would have to be analysed carefully using the fundamental perturbations, as the master field $H$ itself is identically zero for these modes.

\subsection{Pole skipping in the lower half-plane}
\label{sec:PoleSkippingGIFormalism}

\paragraph{}Just like above, we will now identify pole skipping points with frequencies in the lower half-plane by identifying points in momentum space where there are additional regular solutions for $H$. However, we will now be considering the case where $\omega^2\ne v_B^2k^2$: this removes the subtleties in the near-horizon expansion that we dealt with carefully above and we essentially can treat $H$ as a scalar field following \cite{Blake:2019otz}, which we now briefly review.

\paragraph{}As the two independent solutions for $H$ are of the form \eqref{eq:HgenericBCs} near the horizon, it is clear that in order for both solutions to be ingoing we require $\omega=\omega_n=-i2\pi Tn$ where $n=1,2,3,\ldots$. However this condition on its own is not sufficient, as we will now see. To find a Taylor series solution for $H$ at the horizon we make the ansatz
\begin{equation}
    H=e^{-i\omega t+ikx}\left(H_0+H_1(r_0-r)+H_2(r_0-r)^2+\ldots\right),
\end{equation}
and substitute this into the equation of motion \eqref{eq:H-EOM}. Expanding this equation order-by-order near the horizon gives recursion relations for the coefficients $H_m$ which can be written as a matrix equation of the form
\begin{equation}
    M(\omega,k^2)\cdot H=\begin{pmatrix}M_{11} & (2\pi T-i\omega) & 0 & 0 & \cdots \\ M_{21} & M_{22} & (4\pi T-i\omega) & 0 & \cdots \\ M_{31} & M_{32} & M_{33} & (6\pi T-i\omega) & \cdots \\ \cdots & \cdots & \cdots & \cdots & \cdots \end{pmatrix}\cdot\begin{pmatrix}H_0 \\ H_1 \\ H_2 \\ \cdots\end{pmatrix}=0,
\end{equation}
where $M_{ij}=M_{ij}(\omega,k^2)$. At a generic frequency we can solve for all coefficients $H_m$ iteratively in terms of $H_0$: there is a single ingoing solution. At the frequencies $\omega_n$ this is not the case: the first $(n-1)$ coefficients of $H$ form a closed system of equations
\begin{equation}
\label{eq:curlyMeqn}
    \mathcal{M}_n(\omega_n,k^2)\cdot\begin{pmatrix}H_0 \\ H_1 \\ \cdots \\ H_{n-1}\end{pmatrix}=0,
\end{equation}
where $\mathcal{M}_n(\omega_n,k^2)$ is the $n\times n$ matrix obtained by keeping the first $n$ rows and columns of $M(\omega_n,k^2)$. At generic values of $k$, equation \eqref{eq:curlyMeqn} has the unique solution where all coefficients up to and including $H_{n-1}$ vanish. In this case, there is still one ingoing solution: $H_n\ne0$ and all higher coefficients can be solved uniquely in terms of $H_n$. Finally we turn to the special case of interest to us: for specific choices of $k$, the matrix $\mathcal{M}_n(\omega_n,k^2)$ is not invertible and so the equation \eqref{eq:curlyMeqn} has a non-trivial solution. Therefore in this case there are in total two regular solutions: one parameterised by $H_0$ and one by $H_n$. In other words, there is an extra ingoing solution for the modes $\omega=\omega_n=-i2\pi Tn$ and $k=k_n$ where
\begin{equation}
\label{eq:det}
    \det\mathcal{M}_n(\omega_n,k_n^2)=0.
\end{equation}
Solving these polynomial equations for $n=,1,2,3,\ldots$ yields the locations of the pole skipping points $(\omega_n,k_n)$.

\paragraph{}To identify the wavenumber for the pole skipping points with frequency $\omega_1$ we need only the first entry of the matrix
\begin{equation}
\begin{aligned}
    M_{11}=&-32 \pi ^3 T^3 k^4+4 \pi   (4Gs)^2 T (2 \pi  T+i \omega ) \left(2 \pi  T \Vpp -i \omega  \V\right)k^2\\
   &+(4Gs)^2 \omega ^2 \left((2 \pi  T+i \omega ) \left(\Vp^2-\V\Vpp\right)+i \omega  \V^2\right),
\end{aligned}
\end{equation}
where the primes on $V$ denote derivatives with respect to $\phi$.
Setting $\omega=\omega_1$ gives the following quadratic equation for the pole skipping wavevectors $k_1^2$ at this frequency
\begin{equation}
\label{eq:S0eqn2}
    \left(k_1^2+\frac{1}{2}(4Gs)^2(\V-\Vpp)\right)^2+\frac{1}{4}(4Gs)^4\left(2\Vp^2-\Vpp^2\right)=0.
\end{equation}
The two solutions are
\begin{equation}
    k_{1\pm}^2=-\frac{1}{2}(4Gs)^2\Vpp\left(\frac{\V}{\Vpp}-1\pm\sqrt{1-\frac{2\Vp^2}{\Vpp^2}}\right).
\end{equation}
In the high temperature limit $\bar{\lambda}\ll1$, $\Vp/\Vpp\sim\bar{\lambda}$ and so we can expand the square root to obtain the two pole skipping wavenumbers
\begin{equation}
\begin{aligned}
\label{eq:k1sols}
    k_{1+}^2&\,=-\frac{1}{2}(4Gs)^2\left(\V-\frac{\Vp^2}{\Vpp}\right)+\ldots,\\
    k_{1-}^2&\,=-\frac{1}{2}(4Gs)^2\left(\V-2\Vpp+\frac{\Vp^2}{\Vpp}\right)+\ldots,
    \end{aligned}
\end{equation}
where $\ldots$ denotes terms of $O(\bar{\lambda}^3)$ and higher.

\paragraph{}Both solutions \eqref{eq:k1sols} depend on the scalar potential near the horizon. But there is an important qualitative difference between them. Keeping only terms up to $O(\bar{\lambda}^2)$, $k_{1+}^2$ is sensitive only to the quadratic term in $V(\phi)$ i.e.~to the scaling dimension $\Delta$ of the operator $\mathcal{O}$. In contrast, $k_{1-}^2$ is also sensitive to the coefficients $V_3$ and $V_4$ in the scalar potential \eqref{eq:scalarPotential}  i.e.~to more detailed properties of the operator $\mathcal{O}$. In other words, the location of one of the pole skipping points at $\omega=-i2\pi T$ is universal while the other is not.

\paragraph{}Using the explicit expression \eqref{eq:Phi0LambdaRelation} for the horizon value of the scalar field $\Phi_0$ and \eqref{eq:entropyCPT} for the entropy density $s$, we find that the wavenumber of the universal pole skipping point at $\omega=-i2\pi T$ is
\begin{equation}
    k_{1+}^2=-(2\pi T)^2\left(1-2\alpha_\Delta\left(\frac{\pi\Delta(\Delta-2)\cot\left(\frac{\pi\Delta}{2}\right)}{4(\Delta-1)}+1\right)\bar{\lambda}^2\right)+\ldots.    
\end{equation}
This is close to the lightcone and agrees exactly with the prediction \eqref{eq:HydroLHPpoleskippingk} of the resummation of universal hydrodynamics. As we are about to see, the simplicity of this answer is in contrast to generic pole skipping wavenumbers. It would be very interesting to see if it can be identified directly from  one of the horizon Einstein equations for the fundamental metric perturbations, in the same way as the pole skipping point in the upper half frequency plane was in Section \ref{sec:UHPpoleskippingholo}. 

\paragraph{}The wavenumber of the non-universal pole skipping point at this frequency is
\begin{equation}
\begin{aligned}
\label{eq:firstScalarPole}
    k_{1-}^2=-(2\pi T)^2 &\,(\Delta -1)^2\Biggl[1+\frac{6 \sqrt{3} \pi ^{\Delta -\frac{1}{2}} \Gamma \left(\frac{\Delta }{2}\right)}{(\Delta -1)^2 \Gamma \left(\frac{\Delta +1}{2}\right)}V_3\lb \\
&\,-2\alpha_{\Delta}\left(1-\frac{3\pi (16V_4+\Delta (\Delta-2))\cot \left(\frac{\pi 
   \Delta }{2}\right)}{4(\Delta-1)^3}\right)\bar{\lambda }^2\Biggr]+\ldots.   
\end{aligned}
\end{equation}
This is not close to the lightcone and so we cannot compare it to our resummation of hydrodynamics. In the conformal limit $\bar{\lambda}=0$ it reduces to the first pole skipping point \eqref{eq:CFTscalarPSpts} of the $\mathcal{O}$ two-point function in the CFT. This is what we anticipated below equation \eqref{eq:CFTscalarPSpts}: at small $\bar{\lambda}$ the stress tensor mixes with $\mathcal{O}$ and so we expect the stress tensor two-point function to inherit the pole skipping points of the form \eqref{eq:CFTscalarPSpts} up to small corrections.\footnote{Holographically, this mixing is reflected in the coupling between the linearised gravitational and matter perturbations when $\Phi'(r)\ne0$.} The small corrections are due to the expected small changes to the dispersion relations of these modes caused by conformal symmetry breaking. The main message of equation \eqref{eq:firstScalarPole} is that these changes are sensitive to more complicated details of the operator $\mathcal{O}$ than just its dimension. 

\paragraph{}We can repeat this procedure for frequencies $\omega_2$, $\omega_3$, etc.~that are successively lower in the complex plane. The expressions for the matrix elements $M_{ij}$, and even $\mathcal{M}_n$, are very lengthy and unilluminating and so we will not show them explicitly. For $n\geq 2$, the equation \eqref{eq:det} that determines the pole skipping wavenumbers is a polynomial equation for $k_n^2$ of order $n$. Solving it explicitly at $\omega=\omega_2$ we obtain
\begin{equation}
\label{eq:secondScalarPole1}
\begin{aligned}
    &\,k_2^2=-(2\pi T)^2 \Delta ^2\Bigg[1+ \frac{6 \sqrt{3} \pi ^{\Delta -\frac{1}{2}}  \Gamma \left(\frac{\Delta }{2}+2\right)}{(\Delta -1) \Delta ^2 \Gamma \left(\frac{\Delta
   +1}{2}\right)}V_3\lb-2\alpha_{\Delta}\Biggl(1- \frac{\pi\cot\left(\frac{\pi  \Delta }{2}\right)}{16 (\Delta -1)^4 \Delta ^2}\times\\
   &\,  \left(9 \left((\Delta -2)^2 \Delta ^2+8\right) V_3^2+2 (\Delta -1)^2 \Delta ^2 \left(3 \Delta ^2-8 \Delta +4+48
   V_4\right)\right)\Biggr)\bar{\lambda }^2\Bigg],
   \end{aligned}
   \end{equation}
and
\begin{equation}
\label{eq:secondScalarPole2}
\begin{aligned}
    &k_2^2=-(2\pi T)^2 (\Delta -2)^2\Bigg[1-\frac{3 \sqrt{3} \pi ^{\Delta -\frac{1}{2}} (\Delta -4)   \Gamma \left(\frac{\Delta }{2}-1\right)}{4 (\Delta -1) \Gamma \left(\frac{\Delta
   +1}{2}\right)}V_3\lb-2\alpha_{\Delta}\Biggl(1+\\
   &\,\frac{\pi  \left(9 \left((\Delta -2)^2 \Delta ^2+8\right) V_3^2+2 (\Delta -2)^2 (\Delta -1)^2 (\Delta  (3 \Delta -4)+48
   V_4)\right)\cot \left(\frac{\pi  \Delta }{2}\right)}{16 (\Delta -2)^2 (\Delta -1)^4}\Biggr)\bar{\lambda }^2\Bigg].
\end{aligned}
\end{equation}
In these equations we have neglected terms of $O(\bar{\lambda}^3)$ and higher. 

\paragraph{}In principle this procedure can be continued to arbitrarily high order, although for $n\geq 5$ the polynomial cannot be solved exactly: the small $\bar{\lambda}$ expansion must be performed first. The specific expressions for $n>2$ are not very revealing and so we instead describe the pattern of results, which is as anticipated below in the discussion below equation \eqref{eq:CFTscalarPSpts}. For $n\geq 2$, the pole skipping points of the stress tensor two-point function are perturbatively close to those of the $\mathcal{O}$ two-point function in the CFT. These are far from the lightcone and so cannot be accessed by our resummation of hydrodynamics. Furthermore, the corrections are sensitive to the scalar potential coefficients $V_3$ and $V_4$ and so depend on more detailed properties of the operator $\mathcal{O}$ than just its dimension. They do not depend on higher coefficients $V_n$: the equations of motion for the field perturbations depend on $V$ only up to its second derivative, and so when expanded to quadratic order in $\bar{\lambda}$ they depend only on terms up to quartic order in $V$. 

\paragraph{}The pole skipping locations provide constraints on the dispersion relations of poles of the two-point function, and so in principle our results for $k_{1-}^2$, $k_2^2$ etc.~give quantitative information on how the breaking of conformal symmetry affects non-hydrodynamic poles  (see \cite{Grozdanov:2023tag} for related work on reconstructing dispersion relations from pole skipping points). However, given the sensitivity of these results to the details of the potential it seems unlikely that in practice they could be used to extract useful information regarding this.

\paragraph{}In summary, we have shown that generic pole skipping points with frequencies in the lower half-plane are far from the lightcone and are located at wavevectors that are sensitive to details of the scalar potential $V(\phi)$. However, there is one special case of a pole skipping point at $\omega=-i2\pi T$ that is located close to the lightcone at high temperatures. The wavevector of this pole skipping point is universal -- it depends only on the dimension of the operator $\mathcal{O}$ -- and agrees with the prediction \eqref{eq:HydroLHPpoleskippingk} of the near-lightcone resummation of hydrodynamics.

\section{Discussion}
\label{sec:discussion}

\paragraph{}We have computed the locations of the pole skipping points in the stress tensor retarded two-point function of holographic (1+1)d QFTs governed by the action \eqref{eq:bulkaction}. As anticipated, we have shown that there is always a pole skipping point at $\omega_*=+i2\pi T$ and wavenumber $k_*^2=-(2\pi T)^2/v_B^2$, where $v_B$ is the butterfly velocity. We have provided an explicit expression for $v_B$ \eqref{eq:hydrovBprediction} to quadratic order in the high temperature expansion $\bar{\lambda}\ll1$, where it depends only on $\Delta$ and no other details of the QFT. In this high temperature limit we have identified a second pole skipping point whose location depends only on $\Delta$. This lies near the lightcone at $\omega_1=-i2\pi T$ and at a wavenumber $k_1^2$ given in equation \eqref{eq:HydroLHPpoleskippingk}. We have identified additional pole skipping points far from the lightcone, whose locations depend on more details of the CFT than just $\Delta$.

\paragraph{}We also explained how to explain the universal results near the lightcone from a QFT calculation, finding exact agreement with the holographic expressions. This is subtle: near the lightcone, naive conformal perturbation theory in $\bar{\lambda}$ for the stress tensor two-point function breaks down and needs to be resummed. Our results follow from the proposal, building on \cite{Davison:2024msq}, that at leading order in the near-lightcone, high temperature limit
\begin{equation}
\label{eq:nearlightconelimitdiscussion}
    \omega\pm k\sim\bar{\lambda}^2\ll1,
\end{equation}
it should be resummed to
\begin{equation}
\label{eq:resummedGdiscussion}
G(\omega,k)\rightarrow\mp\frac{ck}{24\pi }\frac{(2\pi T)^2+k^2+\ldots}{\omega\mp k\left(1+\Gamma_{\pm}(k)\right)+\ldots}, 
\end{equation}
where
\begin{equation}
    \Gamma_\pm(k)=-\bar{\lambda}^2\frac{\Delta(2-\Delta)}{2(1-\Delta)}\frac{\alpha_\Delta}{1+\left(\frac{k}{2\pi T}\right)^2}\left(\frac{\Gamma\left(2-\frac{\Delta}{2}\right)\Gamma\left(\frac{\Delta}{2}\mp\frac{ik}{2\pi T}\right)}{\Gamma\left(1+\frac{\Delta}{2}\right)\Gamma\left(1-\frac{\Delta}{2}\mp\frac{ik}{2\pi T}\right)}-1\right).
\end{equation}
This resummed expression captures the emergence of hydrodynamics from the microscopic CFT: expanding it for small $\bar{\lambda}$ reproduces the first order conformal perturbation theory result near the lightcone, while expanding it for small $k$ yields the near-lightcone limit of the universal hydrodynamics of \cite{Davison:2024msq}. 

\paragraph{}The very non-trivial agreement between these two calculations suggests that the high temperature, near-lightcone limit of stress tensor dynamics in (1+1)d QFTs is indeed universal, at least in holographic theories. It would certainly be worthwhile to verify the full result  \eqref{eq:resummedGdiscussion} directly for such theories. In higher-dimensional holographic CFTs, the study of thermal correlators near the lightcone has been very fruitful \cite{Fitzpatrick:2019zqz,Kulaxizi:2019tkd,Karlsson:2019dbd,Karlsson:2020ghx,Parnachev:2020fna,Parnachev:2020zbr,Karlsson:2022osn,Huang:2022vet,Esper:2023jeq}. The limit \eqref{eq:nearlightconelimitdiscussion} is particularly physically interesting in (1+1)d for kinematical reasons. Since in (1+1)d the early time CFT excitations and the late time hydrodynamic excitations propagate close to the lightcone, and the butterfly velocity is close to $1$, this limit directly probes features of operator scrambling and the emergence of hydrodynamics: in addition to correctly predicting $v_B$, the expression \eqref{eq:resummedGdiscussion} predicts the dispersion relations of hydrodynamic modes to all orders in the derivative expansion.\footnote{In terms of transport coefficients, it predicts the high temperature limit of all $\Omega_n$ to be the universal expressions of \cite{Davison:2024msq}.} Indeed, \eqref{eq:resummedGdiscussion} essentially says that at high temperatures there is a long-lived excitation near the lightcone at all times. An effective theory for this degree of freedom would be an example of the quantum hydrodynamical effective theory of scrambling proposed in \cite{Blake:2017ris}.

\paragraph{}It would even be beneficial to determine the stress tensor thermal two-point function beyond leading order in the high temperature, near-lightcone expansion. First, evaluating the correction to the numerator of \eqref{eq:resummedGdiscussion} would allow us to explicitly determine the pole skipping frequencies as well as the wavenumbers. Second, evaluating such corrections at small wavenumbers would give access to the remaining hydrodynamic transport coefficients $\kappa_{n,m}$ that enter only in the numerator of the stress tensor two-point function. Third, we expect higher order corrections to resolve the unphysical poles in the dispersion relations of the leading order result \eqref{eq:resummedGdiscussion}. As discussed in Section \ref{sec:hydroreview}, we expect that these apparent poles in fact indicate there are other excitations of the two-point function which approach the lightcone at these isolated (imaginary) wavenumbers, and that the corrections will resolve the apparent poles into branch points.

\paragraph{}It would also be worthwhile to extend our analysis to holographic QFTs theories in which an IR CFT is deformed by an irrelevant scalar with dimension $2<\Delta<3$. The universal hydrodynamics of \cite{Davison:2024msq} is proposed to emerge at low temperatures in these cases and so it seems likely that the stress tensor two point function of these theories in the near-lightcone, low-temperature limit is also given by \eqref{eq:resummedGdiscussion}, at times beyond those where the IR CFT controls the dynamics. For $\Delta>3$ the effects of the $T\bar{T}$ deformation are more important at low temperatures \cite{Delacretaz:2021ufg} and it would be interesting to determine if there is still a simplification in the near-lightcone, low-temperature limit in such cases.
 
\paragraph{}A more challenging question is whether  \eqref{eq:resummedGdiscussion} is valid even for non-holographic QFTs with large $c$. In \cite{Davison:2024msq} it was argued that, when viewed from conformal perturbation theory, the universality of the hydrodynamics that emerges at late times and high temperatures can be traced back to the dominance of stress tensor exchange near the lightcone and an explicit resummation of these effects may be possible. An alternative approach to deriving \eqref{eq:resummedGdiscussion} would be to use the memory matrix formalism of QFT \cite{forster,Hartnoll:2016apf} to naturally isolate and compute the slow relaxation rate $\Gamma_\pm$ of near-lightcone modes at high temperatures. Hamiltonian truncation methods are capable of numerically accessing thermalisation and chaotic dynamics in (1+1)d QFTs \cite{Delacretaz:2022ojg} and so it is plausible that they could also be used to answer this question.

\paragraph{}A technically simpler task that could shed light on this question is to determine the $O(\bar{\lambda}^2)$ correction to the butterfly velocity by computing out-of-time-ordered correlators directly in conformal perturbation theory. In a (1+1)d CFT the butterfly velocity can be obtained from the identity block contribution to the four-point function \cite{Roberts:2014ifa} (see also \cite{Mezei:2019dfv}) and the corresponding contributions to the six-point function are also known \cite{Anous:2020vtw}. Note though that this is not necessarily a direct test of the general validity of the expression \eqref{eq:resummedGdiscussion} for the stress tensor correlator. Firstly, it is not clear if the relation between $v_B$ and the pole skipping wavenumber $k_*$ of this correlator will be true for non-holographic theories. Second, recall that the leading order result \eqref{eq:resummedGdiscussion} does not, on its own, predict the pole skipping wavenumbers. To obtain these without going to the next order, we supplemented \eqref{eq:resummedGdiscussion} with the assumption that the pole skipping frequencies lie at integer multiples of $i2\pi T$, but this is only known to be true in general in holographic theories. Doing this calculation would help to resolve these two important questions (see \cite{Choi:2020tdj} for a study of pole skipping in a non-holographic SYK-like model).

\paragraph{}More generally, there has recently been numerous advances in constraining the thermal correlators of CFTs in (2+1)d and higher using conformal symmetry (see e.g. \cite{Iliesiu:2018fao,Parisini:2022wkb,Marchetto:2023xap,Parisini:2023nbd,Buric:2025anb,Barrat:2025nvu,Buric:2025fye,Niarchos:2025cdg,Barrat:2025twb}). It would be very interesting to see if such methods could be adapted to (1+1)d QFTs at high temperature. In these cases the breaking of conformal symmetry ensures that the QFT shares important physical features with higher dimensional CFTs (e.g.~the emergence of dissipative hydrodynamics) while in some respects remaining simpler, as our results demonstrate.

\acknowledgments
We thank M\'ark Mezei for many helpful conversations and for collaboration in the early stages of this project, as well as for comments on a draft of this manuscript. We are also grateful to Mike Blake, Luca Delacr\'{e}taz, Pavel Kovtun, Andrei Parnachev, Vito Pellizzani, David Ramirez, Andrei Starinets, Petar Tadi\'{c}, Cl\'{e}ment Virally, and Zihan Yan for valuable discussions. We used Sotirios Bonanos' \texttt{RGTC} package and Thomas Hartman's \texttt{GREATER2} package to help with the computations in Section \ref{sec:PoleSkipping}. The work of RD was partially supported by
the STFC Ernest Rutherford Grant ST/R004455/1.  HJ is partially supported by Lady Margaret Hall, University of Oxford.

\appendix

\section{Review of conformal perturbation theory results}
\label{app:comparison}

\paragraph{}In this Appendix we first briefly review the conformal perturbation theory results for the thermal stress tensor two-point function derived in \cite{Davison:2024msq}. Consistency with these was one of the conditions used to derive the universal hydrodynamics at high temperatures, and so they underpin our results in Section \ref{sec:hydroreview} above. We then take the limits $\Delta=1,2$ of these results for general $\Delta$ and compare them with the recent analysis of conformal perturbation theory and pole skipping in \cite{Asplund:2025nkw}.

\subsection{Leading correction for general $\Delta$}

\paragraph{}In the traditional approach to conformal perturbation theory, the first correction to the stress tensor-two point function is proportional to the integrated CFT four-point function $\int dxdy\langle TTO(x)O(y)\rangle$. However, if one is ultimately interested in the momentum space two-point function then there is a slicker way to obtain the correction than evaluating this integral and then Fourier transforming it. We now briefly review this. Further details, and our conventions for retarded two-point functions, can be found in \cite{Davison:2024msq}.

\paragraph{}After coupling to a metric $g_{\mu\nu}(t,x)$ and scalar source $J(t,x)$, the diffeomorphism Ward identity is
\begin{equation}
\label{eq:diffeomorphismWard}
    	\nabla_\mu\langle T^{\mu\nu}\rangle = \sqrt{c}\langle\mathcal{O}\rangle \nabla^\nu J.
\end{equation}
Taking further variations of this with respect to $g_{\mu\nu}$, and then restricting to the flat metric and a constant $J$, gives relations between the stress tensor two-point functions we study. In (1+1)d we find that upon solving these there is actually only one independent two-point function. For example all stress tensor two-point functions in momentum space are related in a simple algebraic way to the two-point function of the trace $ T^{\mu}_{\;\;\mu}$
\begin{equation}
\begin{aligned}
\label{eq:explicitWardidentities1}
    G_{tttt}(\omega,k)&\,=\langle T^{tt}\rangle-\frac{\left(\langle T^{tt}\rangle+\langle T^{xx}\rangle\right)k^2}{\omega^2-k^2}+\frac{k^4}{(\omega^2-k^2)^2}G_{\text{trace}}(\omega,k),\\
    G_{ttxx}(\omega,k)&\,=\langle T^{tt}\rangle-\frac{\left(\langle T^{tt}\rangle+\langle T^{xx}\rangle\right)\omega^2}{\omega^2-k^2}+\frac{k^2\omega^2}{(\omega^2-k^2)^2}G_{\text{trace}}(\omega,k),\\
    G_{tttx}(\omega,k)&\,=-\frac{\left(\langle T^{tt}\rangle+\langle T^{xx}\rangle\right)\omega k}{\omega^2-k^2}+\frac{\omega k^3}{(\omega^2-k^2)^2}G_{\text{trace}}(\omega,k),\\
    G_{xxxx}(\omega,k)&\,=-\langle T^{xx}\rangle-\frac{\left(\langle T^{tt}\rangle+\langle T^{xx}\rangle\right)\omega^2}{\omega^2-k^2}+\frac{\omega^4}{(\omega^2-k^2)^2}G_{\text{trace}}(\omega,k),\\
    G_{xxtx}(\omega,k)&\,=-\frac{\left(\langle T^{tt}\rangle+\langle T^{xx}\rangle\right)\omega k}{\omega^2-k^2}+\frac{\omega^3k}{(\omega^2-k^2)^2}G_{\text{trace}}(\omega,k),\\
    G_{txtx}(\omega,k)&\,=\langle T^{xx}\rangle-\frac{\left(\langle T^{tt}\rangle+\langle T^{xx}\rangle\right)\omega^2}{\omega^2-k^2}+\frac{k^2\omega^2}{(\omega^2-k^2)^2}G_{\text{trace}}(\omega,k),
\end{aligned}
\end{equation}
where we denote the thermal retarded two-point function of $T^{\mu\nu}$ with $T^{\rho\sigma}$ as $G_{\mu\nu\rho\sigma}$. Furthermore, varying \eqref{eq:diffeomorphismWard} with respect to $J$ gives relations between $G_{\mu\nu\rho\sigma}$ and the thermal two point functions of $T^{\mu\nu}$ with $\mathcal{O}$.

\paragraph{}In addition, there is the Weyl invariance Ward identity\footnote{Depending on the value of $\Delta$, there may be extra matter anomalies on the right hand side \cite{Petkou:1999fv} and we will return to these in the next Section.}
\begin{equation}
\label{eq:WeylWard}
    \langle T^{\mu}_{\;\;\mu}\rangle = \sqrt{c}(2-\Delta)J\langle\mathcal{O}\rangle + \frac{c}{24\pi}R,
\end{equation}
where $R$ is the Ricci scalar of $g_{\mu\nu}$. Varying \eqref{eq:WeylWard} with respect to $J$ relates the scalar two-point function $G_{\mathcal{O}\mathcal{O}}$ to the mixed two-point function of $T^{\mu}_{\;\;\mu}$ with $\mathcal{O}$. Combining these with the relations mentioned after equation \eqref{eq:explicitWardidentities1} gives the algebraic momentum space relation
\begin{equation}
\label{eq:explicitWardidentities2}
    G_{\text{trace}}(\omega,k)=-\frac{c}{12\pi}(\omega^2-k^2) + c\lambda^2(2-\Delta)^2\left(G_{\mathcal{O}\mathcal{O}}(\omega,k)-\frac{\Delta}{(2-\Delta)}\frac{\langle\mathcal{O}\rangle}{\sqrt{c}\lambda}\right).
\end{equation}
Combining \eqref{eq:explicitWardidentities1} with \eqref{eq:explicitWardidentities2} gives algebraic relations for any thermal stress tensor retarded two-point function in terms of $G_{\mathcal{O}\mathcal{O}}(\omega,k)$.

\paragraph{}By expressing the stress tensor two-point functions in terms of $G_{\mathcal{O}\mathcal{O}}(\omega,k)$, it is simple to compute the $O(\bar{\lambda^2})$ correction to the CFT result for $G_{\mu\nu\rho\sigma}(\omega,k)$. Due to the $\lambda^2$ prefactor in \eqref{eq:explicitWardidentities2}, to this order we can simply evaluate $G_{\mathcal{O}\mathcal{O}}(\omega,k)$ in the CFT, where it is given explicitly by \eqref{eq:OOCFT}. In other words, the leading correction to the stress tensor two-point function in conformal perturbation theory is controlled just by the scalar two-point function in the CFT. For example, in the notation used in \eqref{eq:CPTstructure} in the main text, this means that the first correction to $G_{tttt}(\omega,k)$ is
\begin{equation}
\begin{aligned}
\label{eq:G2correctionapp}
    G_2(\omega,k)=&\,c (\Delta -2)^2 T^{4-2 \Delta }\frac{ k^4}{\left(\omega^2-k^2\right)^2}G_{\mathcal{O}\mathcal{O}}^{\text{CFT}}(\omega,k)\\
    &+\frac{\pi c \alpha_\Delta T^2}{3}\left(\frac{2\Delta-3}{2(1-\Delta)}+\frac{k^2}{\omega ^2-k^2}-\frac{\Delta(\Delta -2)  }{(\Delta -1) }\frac{k^4}{\left(\omega
   ^2-k^2\right)^2}\right),
   \end{aligned}
   \end{equation}
where we have used the expressions~\cite{Davison:2024msq}
\begin{equation}
\begin{aligned}
    &\,\langle T^{tt}\rangle=\frac{\pi c T^2}{6}\left(1+\alpha_\Delta\frac{2\Delta-3}{\Delta-1}\bar{\lambda}^2+\ldots\right),\quad \frac{\langle\mathcal{O}\rangle}{\sqrt{c}\lambda}=\frac{\pi c T^{2(\Delta-1)}}{3}\frac{\alpha_\Delta}{1-\Delta}+\ldots,\\ &\,\langle T^{xx}\rangle=\frac{\pi c T^2}{6}\left(1+\alpha_\Delta\frac{1}{\Delta-1}\bar{\lambda}^2+\ldots\right),\\
    \end{aligned}
\end{equation}
that are straightforward to obtain using Euclidean conformal perturbation theory on the cylinder \cite{Delacretaz:2021ufg,Davison:2024msq}. Using the expression \eqref{eq:OOCFT} for $G_{\mathcal{O}\mathcal{O}}^{\text{CFT}}(\omega,k)$, more explicitly this is
\begin{equation}
\begin{aligned}
   G_2(\omega,k)=&\,\frac{\pi c T^2  \alpha_\Delta}{3}\Biggl(\frac{2\Delta-3}{2(1-\Delta)}+\frac{k^2}{\omega ^2-k^2}+\frac{k^4}{\left(\omega
   ^2-k^2\right)^2}\frac{(2-\Delta)^2}{(1-\Delta)}\times\\
   &\,\Biggl(\frac{\Gamma\left(1-\frac{\Delta}{2}\right)^2\Gamma\left(\frac{\Delta}{2}-\frac{i(\omega+k)}{4\pi T}\right)\Gamma\left(\frac{\Delta}{2}-\frac{i(\omega-k)}{4\pi T}\right)}{\Gamma\left(\frac{\Delta}{2}\right)^2\Gamma\left(1-\frac{\Delta}{2}-\frac{i(\omega+k)}{4\pi T}\right)\Gamma\left(1-\frac{\Delta}{2}-\frac{i(\omega-k)}{4\pi T}\right)}-\frac{\Delta  }{(2-\Delta) }\Biggr)\Biggr).
\end{aligned}
\end{equation}
Similarly, defining the holomorphic part of the stress tensor as $T=2\pi T_{zz}$ (where $z=x-t$), its thermal retarded two-point function is
\begin{equation}
\begin{aligned}
\label{eq:ourTTCFT}
    G_{TT}(\omega,k)&\,=-\frac{\pi c}{48}\frac{(\omega+k)}{(\omega-k)}\left(\left(\omega+k\right)^2+\left(4\pi T\right)^2\right)+\bar{\lambda}^2G_{TT,2}(\omega,k)+\ldots,
\end{aligned}
\end{equation}
where the first correction to the CFT result is
\begin{equation}
    \begin{aligned}
    \label{eq:ourTTcorrection}
        G_{TT,2}(\omega,k)&\,=\frac{\pi^3 c T^2\alpha_\Delta}{12}\Biggl(4\frac{\omega+k}{\omega-k}+\left(\frac{\omega+k}{\omega-k}\right)^2\frac{(2-\Delta)^2}{(1-\Delta)}\times\\
        &\,\Biggl(\frac{\Gamma\left(1-\frac{\Delta}{2}\right)^2\Gamma\left(\frac{\Delta}{2}-\frac{i(\omega+k)}{4\pi T}\right)\Gamma\left(\frac{\Delta}{2}-\frac{i(\omega-k)}{4\pi T}\right)}{\Gamma\left(\frac{\Delta}{2}\right)^2\Gamma\left(1-\frac{\Delta}{2}-\frac{i(\omega+k)}{4\pi T}\right)\Gamma\left(1-\frac{\Delta}{2}-\frac{i(\omega-k)}{4\pi T}\right)}-\frac{\Delta  }{(2-\Delta) }\Biggr)\Biggr).
    \end{aligned}
\end{equation}

\subsection{$\Delta=1,2$ and comparison with other work}

\paragraph{}We will now compare these results of \cite{Davison:2024msq}, as well as our pole skipping results, to those of \cite{Asplund:2025nkw} for the cases $\Delta\rightarrow1$ and $\Delta\rightarrow2$. 

\paragraph{}The result \eqref{eq:ourTTcorrection} for the correction to the holomorphic stress tensor two-point function has a smooth limit as $\Delta\rightarrow2$ 
\begin{equation}
\begin{aligned}
\label{eq:ourTTcorrectionDel2}
\lim_{\Delta\rightarrow2}G_{TT}(\omega,k)=\frac{\pi^3c}{16}\frac{\omega+k}{\omega-k}\left(\left(\omega+k\right)^2+(4\pi T)^2\right)(2-\Delta)+O((2-\Delta)^2),
\end{aligned}    
\end{equation}
where it vanishes as the operator becomes marginal. It also has a smooth non-vanishing limit as $\Delta\rightarrow1$
\begin{equation}
\begin{aligned}
\label{eq:ourTTcorrectionDel1}
\lim_{\Delta\rightarrow1}G_{TT,2}(\omega,k)=\frac{\pi^3 cT^2}{4}\frac{\omega+k}{\omega-k}\Biggl(4-\frac{\omega+k}{\omega-k}\Bigl(&\,-2+\log{16}+H_{-\frac{1}{2}-\frac{i(\omega-k)}{4\pi T}}\\
&\,+H_{-\frac{1}{2}-\frac{i(\omega+k)}{4\pi T}}\Bigr)\Biggr).
\end{aligned}
\end{equation}
However, the results in these particular limits should be taken with a degree of caution. The Weyl invariance Ward identity \eqref{eq:WeylWard}, from which this result follows, can have extra matter anomalies -- terms proportional to powers of $J$ -- on the right hand side. Typically these appear at higher order in the coupling and so do not affect the corrections that we calculate. But when $\Delta=1$ and $\Delta=2$ there is an anomaly term $J^2$ that should be taken into account \cite{Petkou:1999fv,deHaro:2000vlm} and so may alter the general result in these particular cases.

\paragraph{}Nevertheless, we can compare the results in these limits to the recent computation of the same object in \cite{Asplund:2025nkw}, done by Fourier transforming an analytically continued CFT four-point function integrated on the cylinder.\footnote{In comparing our results with \cite{Asplund:2025nkw} we have taken into account an overall factor of $-T^{2(2-\Delta)}$. The minus sign is from a difference in convention, and the overall scale is because our $G_{TT,2}$ is the correction in units of $\bar{\lambda}^2$ whereas that in \cite{Asplund:2025nkw} is in units of $\lambda^2$. \label{footnote:convention}} In both cases our results are different. In the case of $\Delta=2$, the difference (our \eqref{eq:ourTTcorrectionDel2} $+$(3.31) in~\cite{Asplund:2025nkw}) is
\begin{equation}
\begin{aligned}
    &\,\Delta G^{\Delta=2}_{TT,2}(\omega,k)=-\frac{\pi^3 c}{48}\Biggl(33\omega^2-33k^2-36\omega k-16\pi T\frac{\pi T(k+11\omega)-3i\omega(\omega+k)}{\omega-k}\\
    &\,+12(\omega+k)^2\left(\log16+H_{-1-\frac{i(\omega-k)}{4\pi T}}+H_{-1-\frac{i(\omega+k)}{4\pi T}}\right)\Biggr)(2-\Delta)+O((2-\Delta)^2).
    \end{aligned}
\end{equation}
In the case of $\Delta=1$, the difference (our \eqref{eq:ourTTcorrectionDel1} $+T^2\times $(3.23) in~\cite{Asplund:2025nkw}) is

\begin{equation}
\begin{aligned}
    \Delta G^{\Delta=1}_{TT,2}(\omega,k)=-\frac{\pi^3 c T^2}{4(\omega-k)}\Biggl(-2(k+5\omega)+3(\omega+k)\Bigl(&\,\log16+H_{-\frac{1}{2}-\frac{i(\omega-k)}{4\pi T}}\\
    &\,+H_{-\frac{1}{2}-\frac{i(\omega+k)}{4\pi T}}\Bigr)\Biggr).
    \end{aligned}
\end{equation}
The first thing to note is that these differences are subleading in the high temperature, near-lightcone limit \eqref{eq:IntroHighTNearLCLimit} that we are mainly interested in. In particular, the most singular ~$(\omega-k)^{-2}$ term in \eqref{eq:ourTTcorrectionDel1} agrees very non-trivially with that of \cite{Asplund:2025nkw}. 

\paragraph{}It would be good to understand where the differences come from. One possibility is our neglection of the matter anomalies mentioned above. There are other possibilities we can see just by looking at the thermal CFT two-point function. Ward identities fix this unambiguously (including all contact terms) to the $\bar{\lambda}=0$ limit of our expression \eqref{eq:ourTTCFT}. However, in \cite{Asplund:2025nkw} a CFT result is quoted that comes from performing a position-space integral \cite{Haehl:2018izb,Ramirez:2020qer} and differs from ours by the contact term $\pi c(7\omega^2+k^2+4\omega k+(4\pi T)^2)/48$. One possible origin of this is simply a different definition of the two-point function: see footnote 2 of \cite{Davison:2024msq}, and \cite{Romatschke:2009ng}. Another is that the position space integrals are being regularised in a way that is inconsistent with the Ward identities: see \cite{Keller:2019yrr} for some discussion of the subtleties of regularising such integrals in conformal perturbation theory.

\paragraph{}Putting these differences aside for now, we can also compare our results for the locations of pole skipping points to those of \cite{Asplund:2025nkw}. First, we emphasise that a resummation of the $\bar{\lambda}$ expansion of conformal perturbation theory is necessary to extract the locations of pole-skipping points. Otherwise, the structure of perturbation theory means that the locations of poles are $\bar{\lambda}$-independent (e.g.~see equation \eqref{eq:CPTstructure}). Although no resummation was explicitly described there, \cite{Asplund:2025nkw} does indeed produce the correct $k_*$ for the cases $\Delta=1,2$. Therefore, we believe that the method of \cite{Asplund:2025nkw} must be implicitly resumming first order conformal perturbation theory in the way we have described near $(\omega_*,k_*)$ in these cases. It would be good to understand this more directly: the method of \cite{Asplund:2025nkw} also results in there being no $\bar{\lambda}^2$ correction to $\omega_*=+i2\pi T$, which we instead had to simply assume due to its sensitivity to higher order corrections in the high-temperature, near-lightcone expansion.

\section{Equations of motion for small amplitude perturbations}
\label{app:perturbationequations}

\paragraph{}In this Appendix we present the linearised field equations that we use in Section \ref{sec:PoleSkipping} to determine the locations of pole skipping points.

\paragraph{}In terms of the ingoing null coordinate \eqref{eq:vcoorddefn}, the equilibrium solution \eqref{eq:equilibriummetric} is
\begin{equation}
ds^2=-D(r)dv^2-2\sqrt{B(r)D(r)}dvdr+C(r)dx^2,\quad\quad\quad\quad\quad\phi=\Phi(r).
\end{equation}
We denote the linearised perturbations of the metric around this state as $\delta g_{MN}(r,v,x)$ and the linearised perturbations of the scalar field as $\delta\phi(r,v,x)$. We work in the gauge $\delta g_{Mr}=0$.

\paragraph{}The classical field equations of the action \eqref{eq:bulkaction} are $E_{MN}=0$ and $E_\phi=0$ where
\begin{equation}
\begin{aligned}
    E_{MN}&\,=R_{MN}-\frac{1}{2}\partial_M\phi\partial_N\phi-\frac{1}{2}g_{MN}\left(R+V(\phi)-\frac{1}{2}\partial_A\phi\partial^A\phi\right),\\
    E_\phi&\,=\partial_M\left(\sqrt{-g}\partial^M\phi\right)+\sqrt{-g}\frac{\partial V}{\partial\phi}.
    \end{aligned}
\end{equation}
We use $\delta E_{MN}$ and $\delta E_\phi$ to denote each equation at linear order in amplitude of the perturbations. To manipulate these into the forms presented explicitly below, we assume that $B(r)$, $C(r)$, $D(r)$ and $\Phi(r)$ satisfy the equilibrium equations of motion \eqref{eq:BGEoMs}.

\paragraph{}In total there are seven equations for the small amplitude perturbations. Four of these are second order in radial derivatives and we take them to be
\begin{equation}
\begin{aligned}
    \delta E_{rr}&\,=-\frac{\sqrt{BD}}{2C}\left[\partial_r\left(\frac{C}{\sqrt{BD}}\partial_r\left( \frac{\delta g_{xx}}{C}\right)\right)+\frac{2C\Phi'}{\sqrt{BD}}\partial_r\delta\phi\right],\\
    \delta E_{xx}&\,=-\frac{1}{2}\sqrt{\frac{C}{BD}}\left[\partial_r\left(\frac{C}{\sqrt{BD}}\partial_r\left(\frac{\delta g_{vv}}{\sqrt{C}}\right)-\sqrt{\frac{CD}{B}}\Phi'\delta\phi\right)+\sqrt{C}\Phi'\partial_v\delta\phi\right],\\
    \delta E_{xr}&\,=-\frac{1}{2\sqrt{C}}\left[\partial_r\left(\frac{C^{3/2}}{\sqrt{BD}}\partial_r\left(\frac{\delta g_{vx}}{C}\right)\right)+\sqrt{C}\Phi'\partial_x\delta\phi\right],\\
    \delta E_{\phi}&\,=\partial_r\left(\sqrt{\frac{CD}{B}}\left(\partial_r\delta\phi-\frac{\Phi'\delta g_{vv}}{D} \right)\right)-\sqrt{C}\partial_v\left(2\partial_r+\frac{C'}{2C}\right)\delta\phi-\frac{\Phi'}{2\sqrt{C}}\partial_v\delta g_{xx}\\
    &\,+\frac{\Phi'}{2}\sqrt{\frac{CD}{B}}\partial_r\left(\frac{\delta g_{xx}}{C}\right)+\frac{\Phi'}{\sqrt{C}}\partial_x\delta g_{vx}+\sqrt{BCD}\left(\left.\frac{d^2V}{d\phi^2}\right|_{\phi=\Phi}+\frac{\partial_x^2}{C}\right)\delta\phi.
\end{aligned}    
\end{equation}

\paragraph{}For the remaining three equations, it is simpler to present the linear combinations that are only first order in radial derivatives
\begin{equation}
\begin{aligned}
\label{eq:EinsteinEqsFirstOrder}
    \frac{2D}{C}\left(\delta E_{xr}-\sqrt{\frac{B}{D}}\delta E_{vx}\right)&\,=-\partial_r\left(\frac{\partial_v\delta g_{vx}}{C}\right)+\frac{1}{\sqrt{C}}\partial_r\left(\frac{\partial_x\delta g_{vv}}{\sqrt{C}}\right)-\frac{D\Phi'}{C}\partial_x\delta\phi,\\
    2\left(\delta E_{vr}-\sqrt{\frac{D}{B}}\delta E_{rr}\right)&\,=\frac{1}{2\sqrt{C}}\partial_r\left(\frac{C'\delta g_{vv}}{\sqrt{BCD}}\right)+\frac{\sqrt{C}D{\Phi'}^2}{B}\partial_r\left(\sqrt{\frac{B}{CD}}\frac{\delta\phi}{\Phi'}\right)\\
    &\,-\frac{\partial_r\partial_x\delta g_{vx}}{C}+\frac{1}{\sqrt{C}}\partial_r\left(\frac{\partial_v\delta g_{xx}}{\sqrt{C}}\right)-\frac{D'}{2\sqrt{BD}}\partial_r\left(\frac{\delta g_{xx}}{C}\right),\\
    2C\left(\delta E_{vr}-\sqrt{\frac{B}{D}}\delta E_{vv}\right)&\,=-\sqrt{C}\partial_r\left(\frac{\partial_v\delta g_{xx}}{\sqrt{C}}\right)+\partial_r\partial_x\delta g_{vx}+\frac{D'}{2D}\partial_v\delta g_{xx}\\
    &\,-\frac{D'}{D}\partial_x\delta g_{vx}+\frac{C'}{2D}\partial_v\delta g_{vv}-C\Phi'\partial_v\delta\phi\\
    &\,+\sqrt{\frac{B}{D}}\left(\partial_v^2\delta g_{xx}+\partial_x^2\delta g_{vv}-2\partial_v\partial_x\delta g_{vx}\right).
\end{aligned}    
\end{equation}
These can be combined with the second order equations above to obtain explicit expressions for $\delta E_{vx}$, $\delta E_{vr}$ and $\delta E_{vv}$.

\paragraph{}Our convention for Fourier transformations is
\begin{equation}
    \delta g_{MN}(r,v,x)=\int\frac{d\omega dk}{(2\pi)^2}e^{-i\omega v+ikx}\delta g_{MN}(r,\omega,k),
\end{equation}
and similarly for the scalar field. We use the same symbol for the field in position and momentum space as it should be clear from the context what is meant in any equation.

\bibliographystyle{JHEP}
\bibliography{biblio.bib}

\end{document}